\documentclass[prb,twocolumn,showpacs,superscriptaddress,floatfix,aps,10pt]{revtex4-1}
\usepackage{times}
\usepackage{latexsym,amsmath,amssymb,euscript}
\usepackage{MnSymbol}
\usepackage{color}
\usepackage{dsfont}
\usepackage{multirow}
\usepackage{graphicx}

\begin{document}
\title{Disorder induced transitions in resonantly driven Floquet Topological Insulators}
\date{\today}
\author{Paraj Titum}
\affiliation{Institute for Quantum Information and Matter, Caltech, Pasadena, CA 91125, USA}
\affiliation{Joint Quantum Institute and Joint Center for Quantum Information and Computer Science,
NIST/University of Maryland, College Park, Maryland 20742, USA}
\author{Netanel H. Lindner}
\affiliation{Physics Department, Technion, 320003 Haifa, Israel}
\author{ Gil Refael }
\affiliation{Institute for Quantum Information and Matter, Caltech, Pasadena, CA 91125, USA}

\begin{abstract}
 We investigate the effects of disorder in Floquet topological insulators (FTIs) occurring in semiconductor quantum wells. Such FTIs are induced by resonantly driving a transition between the valence and conduction band. We show that when disorder is added, the topological nature of such FTIs persists as long as there is a mobility gap at the resonant quasi-energy.  For strong enough disorder, this gap closes and all the states become localized as the system undergoes a transition to a trivial insulator. Interestingly, the effects of disorder are not necessarily adverse: we show that in the same quantum well, disorder can also induce a transition from a trivial to a topological system, thereby establishing a Floquet Topological Anderson Insulator (FTAI).  We identify the conditions on the driving field necessary for observing such a transition.
 \end{abstract}
 \maketitle
\section{Introduction}
 Application of a time-periodic drive provides a versatile tool for inducing topological phenomena in condensed matter systems~\cite{Lindner2011,Oka2009,Kitagawa2011,Jiang2011,Lindner2013,Podolsky2013,TorresPRB2014,Gu11,LiuPRL2013,KunduPRB2014}. In addition to  phenomena akin to topological phases in static systems, periodically driven systems exhibit unique topological phases that are only possible in the time dependent setting, \cite{Rudner2013,Titumarxiv2015,Else_bauer_Nayak_PRL2016,Else_Bauer2016,Po_Vishwanath2016,Po_Vishwanath2017,Harper_Roy2016,KhemaniPRL2016} and recent works have focused on classifying such phases\cite{Nathan_NJP2015,Nathan_arxiv2016,Roy_Harper_Prb2016,Else_Nayak_PRB2016,Keyserlingk_PRB2016,Keyserlingk_PRB2016-II,Keyserlingk_PRB2016-III,Potter_PRX2016,Roy_Harper2016,Potter_Morimoto_2016}. Several experiments have probed Floquet topological phase in solid-state\cite{Wang2013,Gedik_Floquet-2}, optical\cite{Rechtsman2013} and cold-atom\cite{Jotzu2014} based systems. In solid state systems, periodic driving applied via external electromagnetic field can induce topological bandstructures for electrons in systems which are topologically trivial absent the drive, thereby yielding ``Floquet topological insulators'' (FTIs). Recent works have explored transport in FTIs,  \cite{TorresPRL2014,Kundu2013,Farrell2015a,Farrell2015b,XiePRB2014}  and means to stabilize topological features in their electronic steady states via coupling to external baths\cite{Seetharam2015,Dehghani2014,Dehghani2014b}. A by and large outstanding question, however, is the extent to which resonantly driven FTIs are robust to disorder.


 In time-independent systems, a topological bandstructure (e.g., bands with non-zero Chern numbers), allows low-dimensional systems to evade localization  as long as they exhibit a bulk mobility gap that separates between the topological bands~\cite{Halperin1982,OnodaPRL2007,ShengPRB2014,Obuse2008,KramerMacKinnon1993,Mirlin2008}. Surprisingly, addition of disorder to a topologically trivial system may  in fact induce a topological phase. Systems which undergo such a transition are referred to as topological Anderson insulators (TAI). Theoretical works showed that it is possible to induce such a transition in a variety of systems, including semiconductor quantum wells~\cite{LiTAIPRL2009,BeenakkerTAI2009}, honeycomb lattices\cite{XingHaldaneTAI2011} as well as in three dimensional semiconductors~\cite{RefaelTAI3D2010}. Experimentally, the transition to a TAI is yet to be observed since the strength of disorder is hard to control in situ.

What are the effects of quenched disorder on topological, periodically driven bandstructures? Could disorder give rise to new topological phases in such systems? These are precisely the questions we consider in this manuscript. We concentrate on the semiconductor quantum well models for Floquet topological insulators\cite{Lindner2011}. In this 2D model, the frequency of the periodic drive is resonant with a transition between the valence and conduction bands, which occurs on a ring in momentum-space. This effectively induces a band-inversion that leads to a topological phase. When disorder is present, and momentum conservation is lost, the band inversion argument cannot be simply applied. In the first part of this work, we demonstrate the robustness of quantum well FTIs when disorder is added, including the persistence of the edge modes in the quasi-energy gap. Additionally, we find that the transition to a trivial phase is similar to the quantum-Hall plateau transition.

In the second part of this work, we study the possibility to induce a transition to a topological phase by adding disorder to a periodically driven semiconductor quantum well. Several previous works have explored the possibility of inducing topological phases by adding disorder to periodically driven systems\cite{Titumarxiv2015,KhemaniPRL2016,Else_bauer_Nayak_PRL2016,Yao_2016,Monroe_2016,Potirniche_2016,Sthitadhi_PRB_2016,Gannot_arxiv2015}. A notable example in a two dimensional system is the Anomalous Floquet Anderson Insulator (AFAI)\cite{Titumarxiv2015}, a phase unique to periodically driven systems which exhibits non-adiabatic quantized pumping. Another possibility was explored in Ref.~\onlinecite{TitumPRL2015}, where it was shown that addition of quenched disorder to a (topologically trivial) periodically driven honeycomb lattice system, induces a transition to a Floquet topological Anderson insulator (FTAI). Such a transition to the FTAI may be realized in an all optical system constructed from helical waveguides in a honeycomb lattice~\cite{Rechtsman2013}.

As we show in the second part of this manuscript, a transition to an FTAI phase can be induced in a periodically driven semiconductor quantum well, thus providing a new route for exploring the FTAI phase in electronic systems. The transition to the FTAI phase induced in the quantum well model relies on a different mechanism than the FTAI induced in the honeycomb system. 
In the honeycomb lattice model, the disorder simply renormalizes the parameters of the undriven effective Hamiltonian in the vicinity of the Dirac points\cite{TitumPRL2015}. In contrast, in the quantum well model, the disorder renormalizes the form of the time-periodic drive. Therefore, even if the form of the drive  induces a topologically trivial Floquet gap in the quantum well absent the disorder, the renormalized drive may produce a topological gap. We find this effect in two variants of the model,  which yield both time reversal symmetric FTAIs, as well as FTAIs with broken time reversal symmetry.
%

The paper is organized as follows. In Sec.~II, we review the model for Floquet topological insulators in a semiconducting quantum well induced by a time-periodic Zeeman field. In Sec.~III, we outline the various  tools used throughout this paper. We discuss how to obtain the localization length from time-evolution of wavepackets, how to probe the localization transition by studying the statistics of the quasi-energy levels, and define the Bott indices for the quasi-energy bands. In Sec.~IV, we demonstrate the robustness of Floquet topological insulators to disorder and the transition to a localized phase at a critical disorder strength. In Sec.~V, we provide a method for realizing the FTAI in the semiconductor model introduced in Sec.~II. Finally, in Sec.~VI, we show the existence of the FTAI phase in a quantum well subjected to an elliptically polarized light.

 \section{Model : Floquet topological insulators}
 We start with a single block of the Bernevig-Hughes-Zhang (BHZ)\cite{BHZ} model of a two-dimensional, spin-orbit coupled, semi-conducting quantum well in the presence of a periodic drive\cite{Lindner2011}
,
 \begin{eqnarray}
H_r({\bf k},t)&=&{\bf d}({\bf k})\cdot \boldsymbol{\sigma} + {\bf V}\cdot \boldsymbol{\sigma}\cos (\Omega t), \label{eq:H_rad}\\
 &\equiv& H_0+V\label{eq:H0+V},
 \end{eqnarray}
where for quasi-momenta around the ${\bf k}=0$ point, we expand ${\bf d}=\left(Ak_x,Ak_y,M-B(k_x^2+k_y^2)\right)$, $k_x$ and $k_y$ being the crystal momentum along $x$ and $y$ directions. The system is driven periodically at a frequency, $\Omega$ with ${\bf V}=(V_x,V_y,V_z)$. The Pauli matrices $\boldsymbol{\sigma}=(\sigma_x,\sigma_y,\sigma_z)$ corresponds to pseudospin. The full BHZ Hamiltonian can be described as a block diagonal $4\times4$ matrix acting on conduction ($E_1$, $\pm1/2$) and heavy hole ($H_1$, $\pm3/2$) states. The Hamiltonian in Eq.~(\ref{eq:H0+V}) will be taken as acting on the two dimensional subspace of positive spins of the conduction ($E_1$, $+1/2$) and heavy hole ($H_1$, $+3/2$) subbands. We label this pseudospin basis as $\{|\uparrow\rangle,|\downarrow \rangle\}$. The periodic drive, ${\bf V}$ may be physically obtained either from a periodic Zeeman field, or an elliptically polarized radiation. While in the presence of a periodically varying Zeeman field, the components, $V_{x,y,z}$ are constants, for an elliptically polarized radiation, ${\bf V}$ is dependent on the crystal momentum, ${\bf k}$. We discuss the details of using elliptically polarized light in section VI and appendix C.

The time-independent part of the Hamiltonian, $H_0$, describes an insulator with spin-orbit coupling. While the full $4\times 4$ BHZ Hamiltonian is time reversal symmetric, the Hamiltonian $H_0$ in Eq.~(\ref{eq:H0+V}) which describes a single block,  is not time-reversal symmetric. Depending on the parameters, it exhibits quantized Hall conductance which is proportional to the Chern number ($C$), a bulk topological invariant. The Hamiltonian is topologically non-trivial when the Chern number is non-zero. The Chern number is $\pm 1$, when $0<M/2B<2$ and, $0$ when $M/2B<0$.
  \begin{figure}
  \includegraphics[width=\linewidth]{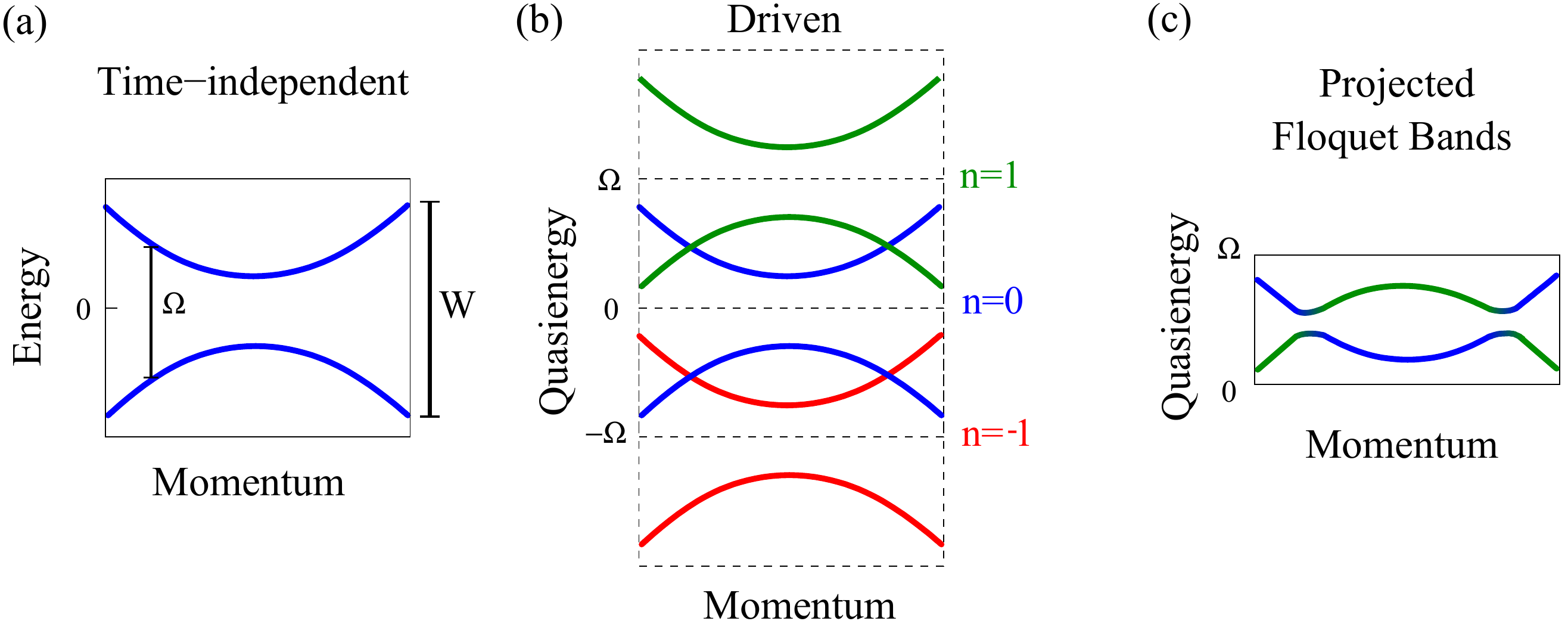}
  \caption{Figure shows the schematic for obtaining the quasi-energies of the driven Hamiltonian. (a) shows example of a time-independent bandstructure in momentum space.  For clarity, we show a slice of the spectrum only along a particular momentum direction. (b) shows three replicas (labeled as $n=-1$, $0$ and $1$, representing three Floquet blocks) of the bandstructure. The replicas are obtained by shifting the time-independent bands by $n\Omega$. This represents the spectrum from the diagonal blocks of $H^F_r$ as shown in Eq. (\ref{eq:FloquetMatrix}). Adjacent Floquet blocks have resonant quasienergies. (c) shows the approximate quasi-energy bandstructure for the Floquet Hamiltonian, obtained by projecting to quasienergies ($\epsilon$) in the range $0<\epsilon<\Omega$. Clearly, the quasienergy bands correspond to  the two resonant bands (Blue and Green) in (b). The radiation potential opens up a gap at the resonance, with the gap proportional to $|{\bf V}_\perp|$ (see Eq. \ref{eq:Vperp}).}
  \label{fig:schematicfloquetfig}
  \end{figure}

 \begin{figure}
  \includegraphics[width=\linewidth]{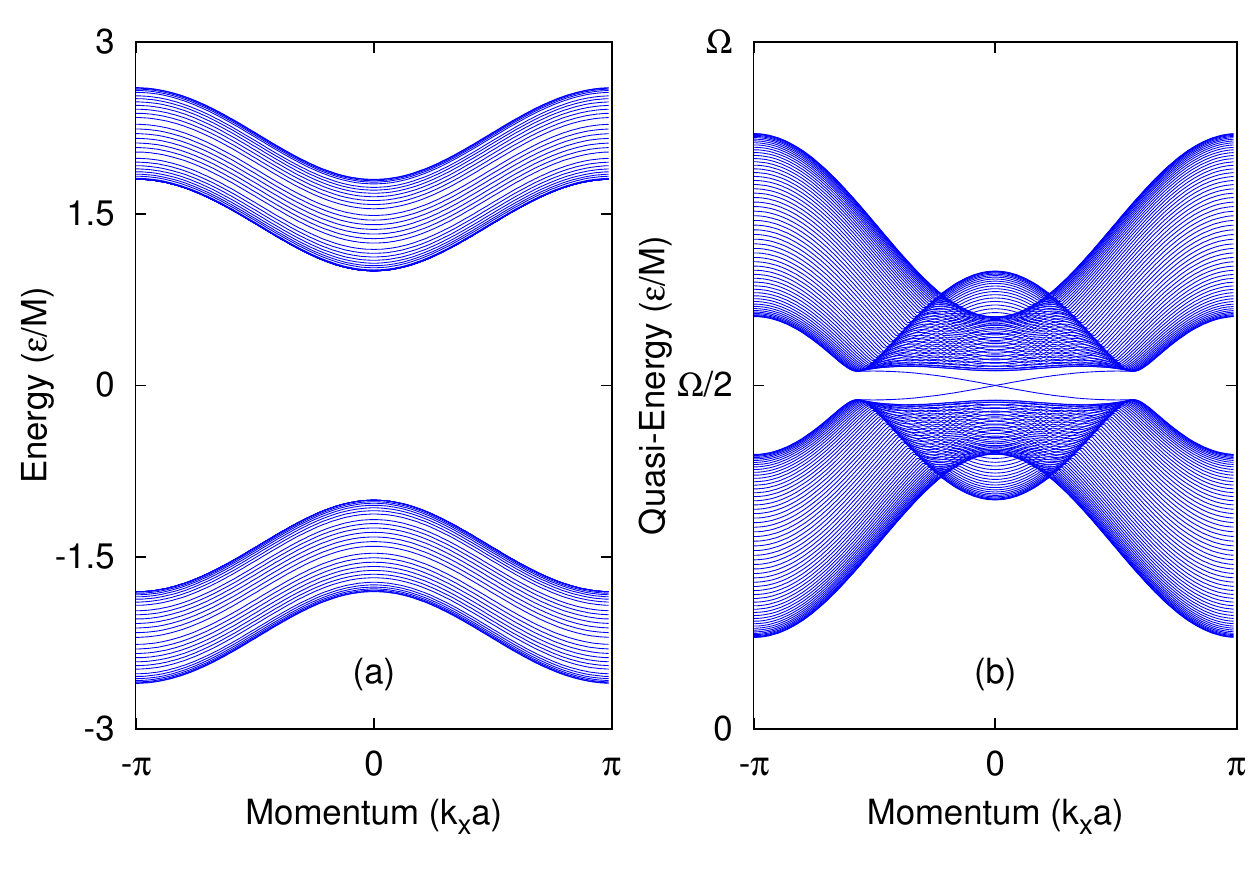}
  \caption{The bandstructure for the model Hamiltonian (defined in Eq. (\ref{eq:H_rad}) ) in different parameter regimes. (a) The original time-independent band-structure for the Hamiltonian for the parameter regimes: $A/M=0.2$, $B/M=-0.2$. The band is topologically trivial. (b) The quasi-energy bandstructure in the presence of driving for the same system parameters as (a), with ${\bf V}/M=(0,0,1)$ and $\Omega/M=3$. This band-structure is clearly non-trivial with edge states in the gap at the resonant quasi-energy, $\epsilon=\Omega/2$. They are done on a lattice with periodic boundary conditions in the $x$ direction and open boundary conditions with $L=60$ sites in the $y$ direction.}
\label{fig:Floquetbandstructure}
 \end{figure}

How can one classify time-dependent Hamiltonians such as the one defined in Eq. (\ref{eq:H_rad})? Energy is no longer a well defined quantum number, however, for a time-periodic Hamiltonian, it is possible to define an effective time-independent Hamiltonian. This so-called Floquet Hamiltonian conserves energy modulo the frequency of the drive, which we label as quasienergy. This is a consequence of the Floquet-Bloch theorem. Periodically driven non-interacting Hamiltonians are analogously classified on the basis of the the topology of the gapped quasi-energy bandstructures. For the cases we study here, Chern and $\mathds{Z}_2$ invariants akin to the invariants used for classifiying static systems are sufficient. More generally, Floquet bandstructures are classified by a winding number\cite{Rudner2013}.

The Floquet Hamiltonian, $H^F_r$ is obtained from a Fourier series in the driving frequency $\Omega$,
 \begin{eqnarray}
  \left(H^F_r\right)_{nm}&\equiv&\langle n|H^F_r({\bf k})|m \rangle \nonumber\\
  &=&n\Omega \delta_{nm}+\int_0^{\frac{2\pi}{\Omega}}dt\ \ H_r(t)\ \ e^{i\Omega(n-m)t}.\label{eq:floquetdef}
 \end{eqnarray}
 Here, $-\infty<n,m<\infty$, are Floquet indices that indicate replicas of the original Hilbert space. Essentially, the periodic time-dependence of the Floquet states, $\{|n\rangle\}$, is determined by the Floquet index, $\langle t|n\rangle=e^{in\Omega t}$. In this formulation, the eigenvalues, $\epsilon$, of $H^F$ are unbounded, but one can identify $\epsilon \rightarrow \epsilon+n\Omega$. Therefore, the quasi-energies of the $H^F$ are periodic consisting of blocks indexed by $n$. Throughout this paper, we use $m$ and $n$ to refer to a given Floquet block. Using Eqs. (\ref{eq:H_rad}) and (\ref{eq:floquetdef}), we note that the time-independent part of the Hamiltonian is diagonal in the Floquet indices while the driving term, acts as hopping between different blocks. Explicitly, the Floquet Hamiltonian has the following form,
  \begin{equation}
   H^F_r=\left( \begin{array}{ccccc}
               \ddots& &\vdots && \\
                \frac{1}{2}{\bf V}\cdot {\boldsymbol \sigma}& {\bf d}({\bf k})\cdot {\boldsymbol \sigma} -\Omega &\frac{1}{2}{\bf V}\cdot {\boldsymbol \sigma} &\udots  \\
                  & \frac{1}{2}{\bf V}\cdot {\boldsymbol \sigma}&{\bf d}({\bf k})\cdot {\boldsymbol \sigma} &\frac{1}{2}{\bf V}\cdot {\boldsymbol \sigma} \\
                  & \udots & \frac{1}{2}{\bf V}\cdot {\boldsymbol \sigma} &{\bf d}({\bf k})\cdot {\boldsymbol \sigma}+\Omega  \\
                  && \vdots   & \ddots\\
              \end{array} \right), \label{eq:FloquetMatrix}
  \end{equation}
 with each row and column corresponding to a particular Floquet index. This matrix formally is infinite dimensional, so, in order to obtain the quasi-energies numerically, we truncate the matrix after $N_F$ blocks, with $N_F$ determined using convergence tests.


 The Floquet topological phase is induced in a trivial BHZ Hamiltonian (i.e. $M/2B<0$) through a resonance in the band-structure. Let us consider the case when the radiation potential induces a single resonance in the bandstructure. This happens when the driving frequency satisfies $\left\{M,\frac{W}{2}\right\}<\Omega<W$, where $W$ is the bandwidth of the time-independent Hamiltonian, $H_0$. The quasi-energy spectrum has a resonance when,
 \begin{equation}
  |{\bf d}({\bf k})|=\Omega/2,\label{eq:resonancecondition}
 \end{equation}
 which corresponds to a circle in the Brilluoin zone. Intuitively, in the perturbative limit, $|{\bf V}|/\Omega\ll 1$, the radiation potential creates a quasi-energy gap at the resonance circle. This can be understood from the schematic diagrams in Fig. \ref{fig:schematicfloquetfig}. Consider a system, $H_0$ with energy bandstrucutre as shown in Fig. \ref{fig:schematicfloquetfig} (a). To obtain the quasi-energy spectrum for the driven system [Eq. (\ref{eq:H_rad})], we first make copies of the bandstructure by translating the original spectrum by $\Omega$ as shown in Fig. \ref{fig:schematicfloquetfig}(b). The resonant quasienergies are modified strongly by the radiation potential, ${\bf V}\cdot {\boldsymbol \sigma}$, resulting in a quasi-energy gap. Then, the quasienergy ($\epsilon$) spectrum is obtained by studying the eigenvalues and eigenstates of $H^F_r$ in one Floquet zone,  $0<\epsilon<\Omega$, as shown in Fig. \ref{fig:schematicfloquetfig} (c).

 Using the rotating wave approximation, the spectral properties of the Floquet Hamiltonian can be approximated by studying an effective two band Hamiltonian of the form\cite{Lindner2011},
  \begin{equation}
H^F_{\rm eff}=\frac{\Omega}{2} \mathds{1}+\left( |{\bf d}({\bf k})|-\frac{\Omega}{2}\right)\hat{d}({\bf k})\cdot{\boldsymbol \sigma} +\frac{1}{2}{\bf V_\perp}.{\boldsymbol \sigma},\label{eq:4}
\end{equation}
where
\begin{equation}
{\bf V_\perp({\bf k})}={\bf V}-\left({\bf V}.\hat{d}({\bf k})\right)\hat{d}({\bf k}). \label{eq:Vperp}
\end{equation}
We denote the quasienergy bands (see Fig. \ref{fig:schematicfloquetfig}) by $\{|\psi^F_\pm({\bf k})\rangle\}$. These bands have a gap at the resonance circle, $V_g=|{\bf V_\perp}({\bf k})|$.

Next, we define the the Chern number $(C)$, which gives the topological invariant for these bands. The Chern number is given in terms of the unit vector, $\hat{n}({\bf k})=\langle \psi^F_-({\bf k})|{\boldsymbol \sigma}|\psi^F_-({\bf k})\rangle$, with
\begin{equation}
 \mathcal{C}=\pm \frac{1}{4\pi}\int d^2{\bf k}\ \ \hat{n}({\bf k})\cdot\left(\partial_{k_x}\hat{n}({\bf k})\times\partial_{k_y}\hat{n}({\bf k})\right),\label{eq:Chernnumberformula}
\end{equation}
where, $k_x$ and $k_y$ are integrated over the first Brilluoin zone. A sufficient condition for the Chern number, $C$, to be non-zero is for the unit vector, $\hat{n}({\bf k})$, on the resonance circle to wind around the north pole~\cite{Lindner2011}. On the resonance circle,
\begin{equation}
\hat{n}({\bf k})={\bf V}_\perp({\bf k})/|{\bf V}_\perp({\bf k})|.\label{eq:nhat}
\end{equation}
  As a consequence, when $\hat{n}({\bf k})$ has zero winding along the resonance circle, the induced phase is trivial. Therefore, in the presence of driving, it is possible to obtain trivial Floquet insulators. For, example, when the radiation potential is along $z$ direction, i.e. ${\bf V}=V_z\hat{z}$, the driven phase is always topological. In contrast, for ${\bf V}=V_x \hat{x}$, the driven system is trivial.

  We obtain the eigenvalues of the Floquet Hamiltonian, defined in Eq.(\ref{eq:FloquetMatrix}), in a cylindrical geometry. The spectrum for an example system is shown in Fig. \ref{fig:Floquetbandstructure}. Fig  \ref{fig:Floquetbandstructure} (a) shows the time-independent bandstructure in the system absent the drive. Fig  \ref{fig:Floquetbandstructure} (b) shows the quasi-energy bandstructure obtained using a resonant drive aligned along $z$ direction, ${\bf V}/M=1\hat{z}$. The presence of chiral edge-modes in the quasi-energy band-gap confirms the non-trivial topology in the driven system.

 We emphasize that so far, we have described the presence of a topological phase in the presence of translation symmetry. In order to study the possible phases in the presence of disorder, we study this model in the presence of quenched disorder on the on-site energies. The disorder potential is chosen to be on-site and $\delta$-correlated,
 \begin{eqnarray}
  V_{\rm dis}=U\mathds{1}, \\
  H=H_r+V_{\rm dis} \label{eq:fullHam}
 \end{eqnarray}
 where $U$ is a uniformly distributed random number between $[-U_0/2,U_0/2]$. The variance of the distribution is $\sigma_d^2=U_0^2/12$.

  The form of the time-periodic Hamiltonians considered in Eq, (\ref{eq:H_rad}) may be realized in several ways in solid-state systems. Let us discuss two possible ways to obtain a non-zero $V$. The first method relies on a Zeeman coupling in the presence of a periodically modulated magnetic field. To be concrete, a magnetic field, ${\bf B}=B_0 \cos(\Omega t)\hat{z}$, results in a driving term, ${\bf V}=\hat{z}(g_E-g_H) \mu_B B_0$, where $g_{E,H}$ is the Zeeman coupling for $E_1$ and $H_1$ pseudospin bands, and $\mu_B$ is the Bohr magneton. Another method to generate this term would be through a periodically modulated  planar electric field, that is introduced through a spatially uniform gauge potential, ${\bf A}=(A_x \sin(\Omega t),A_y \cos(\Omega t))$. In Appendix C we provide details of this model. In this case, ${\bf V}$ is dependent on the momentum vector, ${\bf k}$. As a consequence, the topological phase has a Chern number, $\mathcal{C}=2$ with co-propagating edge-modes.

Finally, we note that within the full four band model, the Floquet bandstucture obtained when the system is subjected to a time-periodic Zeeman field is time-reversal symmetric. Therefore, a topological Floquet spectrum induced by the drive is classified by the same invariant used in the quantum spin hall effect \cite{Lindner2011}. When subjected to a circularly polarized light, the Floquet spectrum of the full four band model is not time reversal symmetric. Therefore in the latter case, it exhibits bands whose topological properties are captured by Chern numbers, which for the present model can take the values $\pm2$, depending on the chirality of the polarization.

\section{Diagnostics for Drive and disorder induced topological transitions}
In this section, we will outline the numerical methods used to analyze the driven-disordered systems. First, we use an approximate real time-evolution of wave packets to obtain the single-particle transport properties in the presence of disorder. Second, we obtain the exact Floquet quasi-energy eigenstates by diagonalizing the stroboscopic time-evolution operator integrated over a single time-period, $T$. The topological nature of these disordered bands are investigated by computing the Bott index\cite{hastingsLoringEPL2010}, which is equivalent to the Chern invariant in the presence of disorder. The localization-delocalization transition is also further examined by computing the eigenvalue statistics to identify the transitions in these systems.
\subsection{Time evolution}
This method is used to numerically obtain the disorder-averaged transmission probability at a particular quasi-energy. For a given disorder realization, we employ a numerically exact time evolution to determine the sample-dependent propagator as a function of disorder and quasi-energy. The sample averaging is then done by repeating the procedure over many realizations.

 We initialize the system with a $\delta$-function wavepacket in real space and study how this wavepacket spreads in real space. The bulk or edge properties are determined according to the choice of the initial position of the wavepacket and boundary conditions. While the bulk Green's function is obtained from the Hamiltonian in a torus geometry with periodic boundary conditions in both the $x$ and $y$ directions, the edge Green's function is obtained in a cylindrical geometry with open boundary conditions along $y$. In our simulations we always initialize with the up pseudospin, at position ${\bf x}$, $|{\bf x},\uparrow\rangle$, where $\sigma_z|\uparrow\rangle=|\uparrow\rangle$. The time evolution operator is  $U(t,0)=T\exp(-i\int_0^t H(t') dt')$, where $H(t)$ is defined in Eqs. (\ref{eq:H_rad}) and (\ref{eq:fullHam}) . The time-evolution operator is obtained numerically using a split operator decomposition. The exact details of the numerical procedure is provided in Appendix B. The time-evolution of a $\delta$-function wavepacket for $N$ time periods, gives us the wavefunction in real space, which is also the propagator,
\begin{eqnarray}
 G({\bf x},{\bf x}',NT)&=&\langle {\bf x}' \uparrow|U(NT,0)|{\bf x}\uparrow\rangle. \label{eq:GF(x,x't)}
\end{eqnarray}
In order to explore the effect of disorder and quasi-energy, we must average the Fourier transform  $\mathcal{F}\left[G({\bf x},{\bf x}',NT)\right]\equiv G_N({\bf x},{\bf x}',\epsilon)$, over a number of disorder configurations. Let us define, ${\bf x}\equiv(x,y)$  ${\bf x}'\equiv (x',y')$. The average transmission probability along $x$ direction as a function of $x'$ is,
\begin{eqnarray}
 g_N(x, x',\epsilon)&=&\overline{\sum_{y'}|G_{N}({\bf x},{\bf x}',\epsilon)|^2}, \label{eq:15}
\end{eqnarray}
where $\overline{ (\cdots)}$ denotes disorder averaging, and the subscript, $N$ is related to the total time of evolution, $T_{\rm tot}=NT$. The properties of $g_N$ are used to identify a localization transition in this system.  We extract the decay length scale, $\Lambda_N(\epsilon)$, of the transmission from the inverse participation ratio of $g_N$. The details of the exact definition is provided in Appendix B. 
The mobility-gap in the spectrum is identified by states satisfying $\Lambda_N(\epsilon)/L \ll 1$, where $L$ is the system size.  Clearly, $\Lambda_N$ depends on the total time of evolution, $T_{\rm tot}$. Therefore, the time-dependence of the length scale, $\Lambda_N(\epsilon)$, is obtained by computing $\Lambda_N$ from Eqs. (\ref{eq:GF(x,x't)}) and (\ref{eq:15}), for different number of time-periods, $N$.
\subsection{Floquet-spectrum}
 The quasi-energy spectrum can be numerically obtained from the unitary time evolution operator, $U(T,0)$ for a single time- period. Let the infinitesimal time-evolution operator for a time step, $\delta t$ be, $U(p\delta t,(p-1)\delta t)$, where $p$ is an integer. The time-evolution operator for a full time-period through a time ordered product,
 \begin{equation}
 U(T,0)=\prod_{p=1}^{N_{\rm div}}U(p\delta t,(p-1)\delta t),  \ \ {\rm with,}\ \ \delta t=T/N_{\rm div}.
 \end{equation}
 We provide the details of the calculation of the time evolution operator in Appendix B. Now, the Floquet quasi-energy spectrum, and eigenvectors are obtained by diagonalizing, $H^F=\frac{i}{T}\ln(U(T,0))$. Here, the branch cut of the logarithm is chosen such that, $-i\ln e^{i\epsilon T}=\epsilon T\ \ \textrm{ for all}\ \ \epsilon T\in [0,2\pi)$ .  For our purposes, it is convenient to rewrite the disorder potential in the momentum space and obtain the eigenvectors in this space. To characterize the topological of the Floquet Hamiltonian, we compute the Chern number of the quasi-energy bands, by calculating the Bott index from its eigenvectors.
\subsubsection{Bott Index}
 The Bott index is a topological invariant that has been defined by Loring and Hastings\cite{hastingsLoringEPL2010} for disordered two dimensional systems with no additional symmetries except for particle number conservation. For time-independent Hamiltonians, it has also been shown that this index is equivalent to the Hall conductivity of the sample, which in turn is given the Chern number\cite{HastingsLoringJmathPhys}.

 We define an analogous Bott index for a time periodic Hamiltonian using the eigenstates of the Floquet Hamiltonian. Consider a projector $P(\epsilon_l,\epsilon_h)$ on a band of Floquet states of a two dimensional system with periodic boundary conditions, where the band is bounded  by $\epsilon_l<\epsilon<\epsilon_h$. We define two unitary matrices, $   U_X= \exp(i2\pi X/L_x)$ and,  $ U_Y= \exp(i2\pi Y/L_y)$, where $X$ and $Y$ are diagonal matrices indicating the $x$ and $y$ coordinate respectively. The Bott index is an integer which measures of commutativity of the projected unitary matrices $\tilde{U}_{X}=PU_{X}P$ and $\tilde{U}_{X}=PU_{X}P$, and is is given by\cite{HastingsLoringJmathPhys}
   \begin{equation}
 C_b(\epsilon_l,\epsilon_h)=\frac{1}{2\pi}{\rm Im}\left[{\rm Tr}\left\{\ln\left(\tilde{U}^{}_{Y}\tilde{U}^{}_{X}\tilde{U}^{\dag}_{Y} \tilde{U}^{\dag}_{X} \right)\right\}\right].
\end{equation}.

 The Bott index $C_b(\epsilon_l,\epsilon_h)$ is an integer. As shown in Ref.~\onlinecite{HastingsLoringJmathPhys}, for time-independent Hamiltonians $C_b$ is equivalent to the quantized Hall conductance obtained by filling the band associated with the projector $P$, and thereby, it is equivalent to the Chern number\cite{Avron1985} associated with the projector $P$. A direct consequence is that the Bott index can be related the difference between the number of chiral edge states above and below the band. In a disk geometry , defining $n_{\mathrm{edge}}( \epsilon_h)$ to be the total number of right moving chiral edge states at $\epsilon_h$ (in a translation invariant system, $n_{\mathrm{edge}}( \epsilon_h)$ is the number of right moving edge states minus the number of left moving edge states at $\epsilon_h$), and similarity defining $n_{\mathrm{edge}}( \epsilon_l)$, the Bott index is related to the number of edge states by~\cite{Rudner2013} $C_b(\epsilon_l,\epsilon_h)=n_{\mathrm{edge}}( \epsilon_h)-n_{\mathrm{edge}}( \epsilon_l)$ .

 Having set $\epsilon_h$ and $\epsilon_l$, the average Bott $\overline{C}_b$ for a given disorder strength is obtained by averaging this index over different disorder configurations. While the index $C_b(\epsilon_l,\epsilon_h)$ is an integer for every particular disorder configuration, there is no such requirement for the average Bott index. For a given system size, upon increasing disorder and crossing a topological phase transition, the average index smoothly changes from one integer to another\cite{hastingsLoringEPL2010}. This smooth transition is expected to be sharp in the thermodynamic limit.

\subsubsection{Level spacing statistics.}
 The statistics of the eigenvalues of the Floquet Hamiltonian can be used to study the localization-delocalization transition\cite{Shklovskii_PRB_1993}. Generically, extended eigenstates experience level repulsion, localized states with vanishing overlap can be arbitrarily close to each other in quasi-energy. This leads to different behaviors in the distribution of the spacings of the quasienergies for regions in the spectrum corresponding to localized and extended states

Below, we outline the characteristic features of the level-spacing distributions for extended and localized states. The level-spacing, $s_n$, is defined as the difference of adjacent quasi-energies, $s_n=\epsilon_{n+1}-\epsilon_n$. Let the average level spacing over all quasienergies and disorder realizations be $\delta$. We now define level-spacing at a given quasi-energy, measured in units of $\delta$, $s(\epsilon)=\frac{1}{\delta}\sum_{|\epsilon_n-\epsilon|\leq d\epsilon}s_n$, where we choose the window of quasi-energy such that $d\epsilon\gg s_n$. The probability distribution of the level-spacings, $P(s)$, is universal and determined completely by their symmetry classification. Given a Hamiltonian with broken time-reversal symmetry, it corresponds to the Unitary class. Extended eigenstates must have level-repulsion and, so, the level spacing distribution at this quasi-energy, $P(s)$, corresponds to the Gaussian Unitary Ensemble (GUE)\cite{Mehta2004},
\begin{equation}
P_{\rm GUE}(s)=\frac{32}{\pi^2}s^2\exp\left(-\frac{4}{\pi}s^2\right).
\end{equation}
For Floquet systems, the correct ensemble corresponds to the Circular Unitary Ensemble (CUE), since the eigenvalues are defined on a compact manifold. The distributions for CUE and GUE are expected to converge to the same distribution in the thermodynamic limit\cite{DAlessio2014}.
In contrast, for localized states, the level-spacings must follow Poisson statistics,
 \begin{equation}
  P_{\rm loc}(s)=\exp(-s).
 \end{equation}
We note that the variance of the GUE distribution $\sigma^2(P_{\rm GUE}(s))=\int s^2 P_{\rm GUE}(s)ds- \left(\int s P_{\rm GUE}(s)ds\right)^2\sim 0.178$ differs from the variance of the Poisson distribution, $\sigma^2(P_{\rm loc}(s))\sim 1$.
 \section{Localization transition in disordered FTIs}
 In this section, we study the effect of adding on-site quenched disorder to the periodically driven system given in Eq.~(\ref{eq:H_rad}). We choose the system parameters such that the Floquet spectrum is topological. In this section, we consider the radiation potential to be of the form ${\bf V}=(0,0,V_z)$, see Eq.~(\ref{eq:H_rad}). In the following, we focus on a single $2\times 2$ block of the full time-reversal invariant $4\times 4$ Hamiltonian. As discussed in Section II, the driving frequency is chosen to induce only a single resonance, such that the Floquet bands have a non-trivial Chern number, $|C|=1$. In the following, we show that
 the topological phase is robust to disorder, with gapless edge states, for disorder strengths $\sigma_d\ll V_g$, where $V_g$ is the quasienergy gap of the clean system. At sufficiently strong disorder, $\sigma_d\gg V_g$, the mobility gap closes and all eigenstates are completely localized.

 \subsection{Analytical discussion of perturbative corrections}
  The correction to the Floquet density of states is obtained perturbatively in the disorder potential. The effect of the disorder potential is to renormalize the parameters of the Hamiltonian. To lowest order in the disorder potential, the self-energy corrections to the single particle Floquet Green's function is obtained in the Born approximation, the details of which are provided in Appendix A. As discussed in Section II, the quasi-energy gap at resonance is determined to lowest order by the radiation potential, ${\bf V}_\perp$ [see Eq. (\ref{eq:Vperp})].    We define the self-energy correction to ${\bf V}$ as $\Sigma_V$. In the Born approximation, we consider the dominant contribution to the self-energy correction to the radiation potential, $\Sigma_V$. These corrections yield a renormalization of the density of states at resonance.

 As we show in detail in appendix A, in the Born approximation $\Sigma_V$ is proportional to $-\overline{U^2} \sigma_z$, where $\overline{U^2}=\sigma_d^2$ is the variance of the distribution of the disorder potential. Therefore, this term renormalizes (negatively) the magnitude of $V$ indicating that the quasienergy gap must close with disorder. Thus qualitatively, the Born approximation captures the transition from topological to trivial insulator as a result of closing of the quasi-energy gap. The Born-approximation, however, fails to capture the transition to an Anderson localized insulator as the gap closes. The details of the Born-approximation calculation are provided in the Appendix A.
 \subsection{Numerical Analysis}
 We study the quasi-energy spectrum numerically. The numerical simulations were done for Hamiltonian with the parameters $A/M=0.2$ and $B/M=-0.2$ on a torus with dimensions $L_x\times L_y$. We now summarize the numerical results, and elaborate on them below. First, we examine the mobility gap in the quasi-energy spectrum: the spectral region around quasienergy $\Omega/2$ containing localized states. The average bulk localization length, $\Lambda_N$ as a function of disorder and quasi-energy is shown in Fig. \ref{fig:2.5} (a). Clearly, with increasing disorder the mobility gap centered at quasi-energy $\Omega/2$ vanishes. Next, we identify the point where the mobility gap closes as a topological-to-trivial phase transition. We calculate the topological invariant for all states below the  resonant quasi-energy. The average value of the invariant changes as a function of disorder, as shown in Fig. \ref{fig:2.5} (b). Finally, we study the level-spacing statistics.  At weak disorder, the level-spacing statistics for states inside the quasi-energy bands corresponds to the value attained for extended states (indicating that the localization length is larger then the system size studied). However, at strong disorder after the mobility gap closes,  the level-spacing statistics of all Floquet eigenstates correspond to localized states. This transition to localized states is shown in Fig. \ref{fig:3}. This transition is in line to a localization transition in the quantum Hall universality class, where the phase transition corresponds to a vanishing mobility gap in the spectrum.
 \subsubsection{Results from bulk localization length.}
 \label{subsubsec: loc length}
  We study the time-evolution of a localized wave-packet under the Hamiltonian defined in Eq. (\ref{eq:fullHam}).   The initial wavepacket is chosen to be a  $\delta$-function in real space, peaked at position ${\bf x}$, and corresponding to a positive eigenvalue of $\sigma_z$. We denote this state by  $|{\bf x} \uparrow\rangle$. The numerical simulations are done with periodic boundary conditions along both $x$ and $y$ directions. This allows us to probe the bulk Floquet states. The average transmission probability from $x$ to $x'$, given by $g_N(x,x',\epsilon)$, is obtained from Eq. (\ref{eq:15}). We define the bulk localization length, $\Lambda_N$, as the inverse participation ratio (IPR) of $g$, the details of which are provided in Appendix B. Fig \ref{fig:2.5} (a) shows $\Lambda_N/(2L_y)$ as a function of quasi-energy and disorder. At small disorder ($U_0/M<0.6$), near the resonance, $\epsilon\sim \Omega/2$, the transmission decays with a localization length $\Lambda_N/L_y \ll 1$. We identify the spectral region around quasi-energy $\Omega/2$ with $\Lambda_N/L_y \ll 1$ as a 'mobility gap'. In contrast, the eigenstates in the bands, for quasi-energies well away from resonance
exhibit a localization length which is on the order or larger then the system size. This is in line with what we expect from quasi-energy bands with non-zero Chern numbers. A band exhibiting a non-zero Chern number necessarily exhibits at least one delocalized state, and a diverging localization length at quasi-energies approaching the quasi-energy of this state. In the finite sizes that we examine numerically, the localization length exceeds the system size for most of the states in the bulk band, and therefore, the entire band seems to be delocalized. 
It is clear that with increasing disorder strength, the mobility gap vanishes. The localization length, $\Lambda_N(\epsilon)$ is also a function of the total time of evolution ($T_{\rm tot}=NT$), as discussed in section III A.  The bulk localization length for a typical sample at a given quasi-energy $\epsilon$, appears to be diffusive for short times, $\Lambda_N^2\sim DN$ where $D$ is the diffusion constant. At long times,the localization length saturates, $\lim_{N\to \infty}\Lambda_N \rightarrow \Lambda_{\rm sat}$, either to the system size or the true localization length at that quasi-energy. We provide numerical evidence for the time-dependence of $\Lambda_N$ in Appendix B.
 \begin{figure}
  \includegraphics[width=\linewidth]{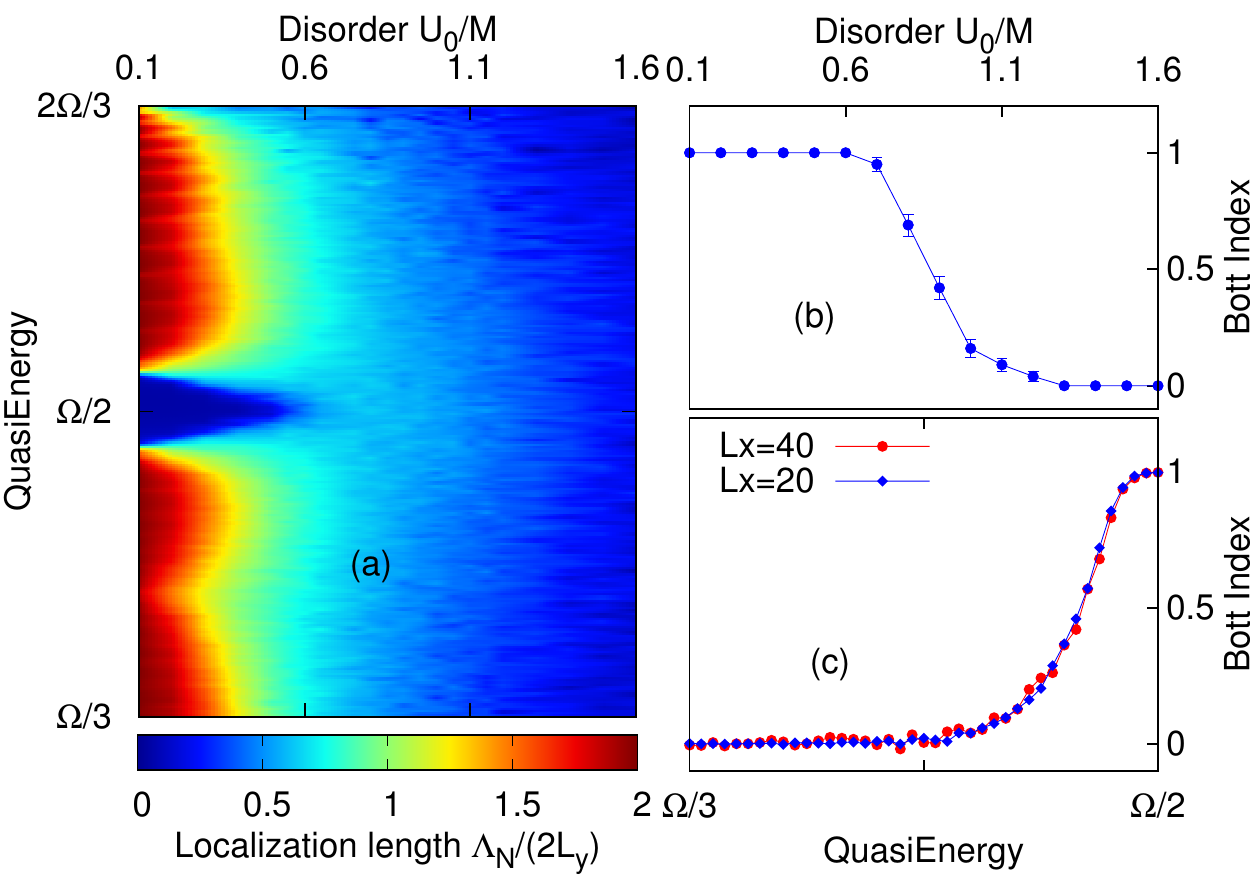}
  \caption{(a)  The average localization length $\Lambda_N/(2L_y)$ as a function of Quasi-energy and disorder strength. This figure shows  transition  in which the mobility gap closes at $U_0/M\approx 1$. Simulations were done for system sizes $L_x\times L_y=400\times50$, and evolved for $N=2500$ time periods. (b) shows the topological phase transition accompanied by the mobility gap closing. We plot the disorder-averaged Bott index, $\overline{C}_b(0,\Omega/2)$ for states between $\epsilon_l=0$ and $\epsilon_h=\Omega/2$. At small disorder the system is topological with a Bott index, $1$. At $U_0/M\approx 1$, the mobility gap closes and the system becomes trivial with the index going to $0$. (c) Disorder averaged Bott index as a function quasienergy, $\overline{C}_b(0,\epsilon)$, for disorder strength, $U_0/M=0.6$. The index takes the value $\overline{C}_b(0,\epsilon)\approx 1$ for  $\epsilon=\Omega/2$ and smoothly goes to zero for quasi-energy $\epsilon$ away from $\epsilon=\Omega/2$. We plot for two different system sizes, $L_x=20$, $40$.}\label{fig:2.5}
 \end{figure}
 \subsubsection{Bott index of Floquet bands.}
 The clean Hamiltonian is topologically non-trivial with a quasi-energy band-structure shown in Fig. \ref{fig:Floquetbandstructure} (b). As long as a mobility gap exists, the topological phase remains robust to disorder. This is confirmed by measuring the  disorder-averaged topological invariant, $\overline{C}_b(\epsilon_l,\epsilon_h)$ (Bott index) of a quasienergy band of states defined by $\epsilon_l<\epsilon<\epsilon_h$, as shown in Fig \ref{fig:2.5} (b). In the numerical computation, we set $\epsilon_l=0$. For weak disorder strength, the average index at the quasi-energy $\epsilon_h=\Omega/2$, is $\overline{C}_b(0,\Omega/2)=1$, indicating a topological phase. After the mobility gap closes the index smoothly goes to zero. This indicates a transition from a topological to a trivial phase which is expected to be sharp in the thermodynamic limit.
In analogy to disordered integer quantum Hall systems, each quasi-energy  band necessarily exhibits a critical quasi-energy whose corresponding (bulk) eigenstate is extended. This extended state carries the non-zero topological invariant of the band. To probe this critical energy, we plot the index as a function of quasi-energy in Fig \ref{fig:2.5}(c). We see a smooth transition from topological to trivial as a function of quasi-energy. This transition as a function of the quasi-energy is expected to happen at the critical quasi-energy. However, the system sizes we consider are not sufficient to reveal the critical energy in the quasi-energy bands.
 \subsubsection{Level-spacing statistics.}
\begin{figure}
  \includegraphics[width=\linewidth]{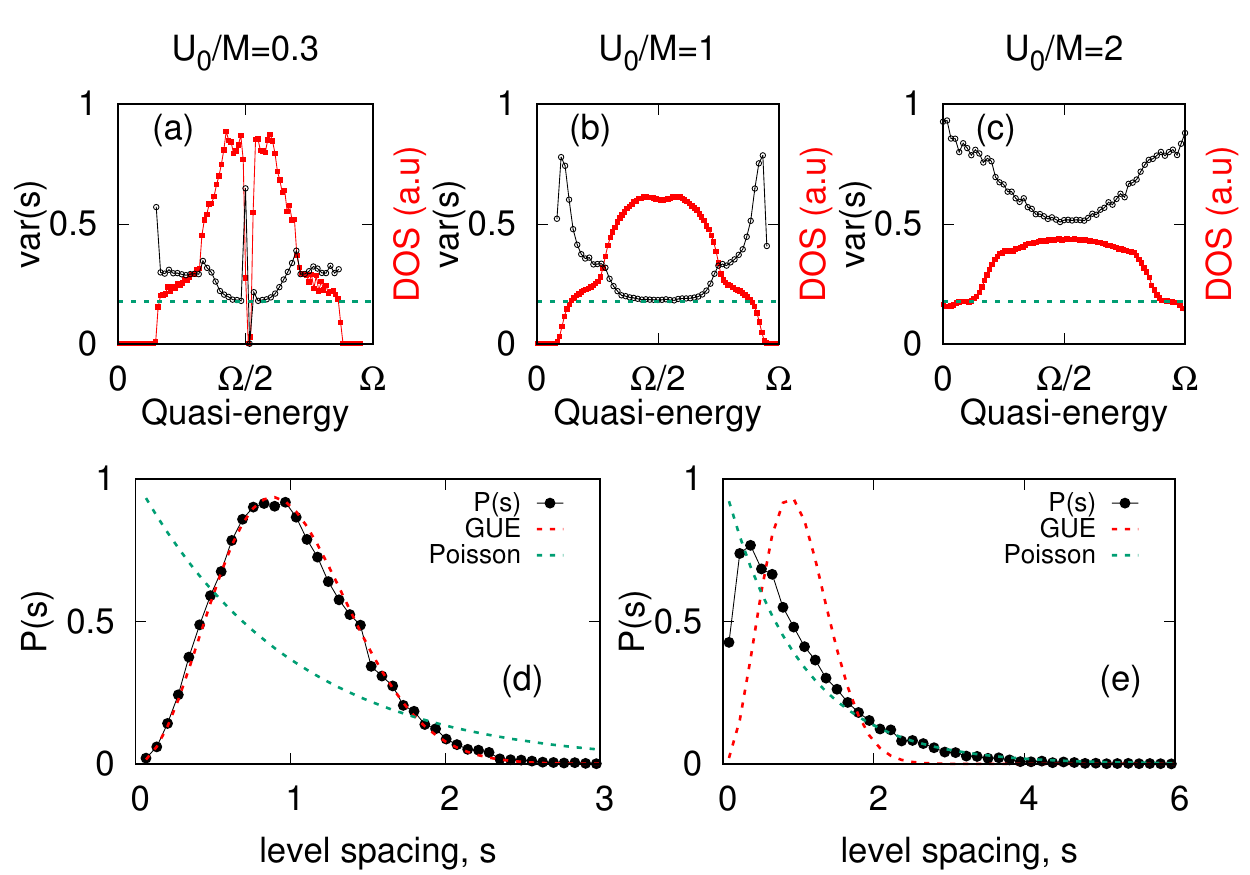}
  \caption{Level spacing statistics for the Floquet eigenvalues at different disorder strengths. The extended Floquet states must have level repulsion, with level statistics following the GUE ensemble, in which case, the variance of the level spacing distribution, $\sigma^2(P(s))=0.178$. This is in contrast to localized states which have $\sigma^2(P(s))\sim 1$. The level spacing $s=\Delta\epsilon/\delta$ is measured in units of average level-spacing, $\delta$, as defined in Section III. Figs. (a), (b) and (c) compare the density of states of the Hamiltonian (in red) with the variance of $P(s)$ (in black) for disorder strengths, $U_0/M=0.3$, $1$, and $2$ respectively. Panel (a) shows data which corresponds to two quasi-energy bands with extended states (indicating that the localization length is larger than the system size). The two bands are separated by a band-gap. In (b) the band gap is no longer visible while the level statistics of states in the middle of the quasi-energy zone still corresponds to extended states, and in (c) all the Floquet eigenstates are localized. (d) shows the level spacing distribution, $P(s)$, at a given quasi-energy, $\epsilon/\Omega=0.43$ and disorder strength $U_0/M=0.3$. It exactly fits with the GUE distribution. (e) shows $P(s)$ for $\epsilon/\Omega=0.25$, at disorder strength $U_0/M=2$. This distribution has better
agreement with poisson statistics, indicating localized states. All the simulations were done for systems sizes $L_x\times L_y=40\times40$.}\label{fig:3}
 \end{figure}

 We analyze the level spacing statistics to probe the nature of the localization transition in these driven Hamiltonians. The Floquet Hamiltonian, $H_F$,  corresponding to a single $2\times 2$ block, as defined in Eq. (\ref{eq:FloquetMatrix}),is not time-reversal symmetric. In the topological phase, the quasi-energy bands have non-zero Chern invariant. We therefore expect that indicates that the localization-delocalization transition must be of the quantum-Hall universality class. We confirm this by studying the level spacing statistics as a function of quasi-energy for different disorder strengths as shown in Fig. \ref{fig:3}. At weak disorder [Fig. \ref{fig:3} (a)], the states have level-spacing statistics identical to the GUE ensemble, $P_{\rm GUE}$ as shown in Fig. \ref{fig:3} (d). This is an indication that the localization length of states in the quasi-energy bands is much larger than the system sizes we studied numerically. This is expected for bands with non-zero Chern numbers in the presence of weak disorder,  as discussed in Sec.~\ref{subsubsec: loc length}. At $U_0/M\approx 1$, the mobility gap is closed as shown in Fig. \ref{fig:3} (b). At much stronger disorder, $U_0/M\approx2$, we notice that the Floquet bands are completely localized and obey Poisson statistics as shown in Fig. \ref{fig:3} (e). This is in agreement with our expectation that the quasi-energy bands exhibit non-zero Chern numbers and therefore delocalized states, which persist as long as the mobility-gap around quasi-energy $\Omega/2$ exists.
\section{Disorder-induced topological phase: The FTAI}
  We now discuss the realization of a disorder-induced topological phase in a driven but topologically trivial quantum-well. This phase is a driven analog to Topological Anderson Insulators (TAI), predicted numerically\cite{LiTAIPRL2009} first in quantum wells. Floquet Topological Anderson insulators (FTAI) are the analogous disorder induced topological phase and have been predicted in the driven honeycomb lattice models\cite{TitumPRL2015}. In the following, we first review the prescription for realizing Topological Anderson insulators in quantum-well structures. Then, we propose a method for realizing an FTAI in the driven quantum well system, as well as numerical evidence supporting the realization of FTAI using our proposal. We note that The FTAI realized in the quantum well model is time-reversal invariant, in contrast to the one realized in the honeycomb lattice models.
  \begin{figure}
  \includegraphics[width=\linewidth]{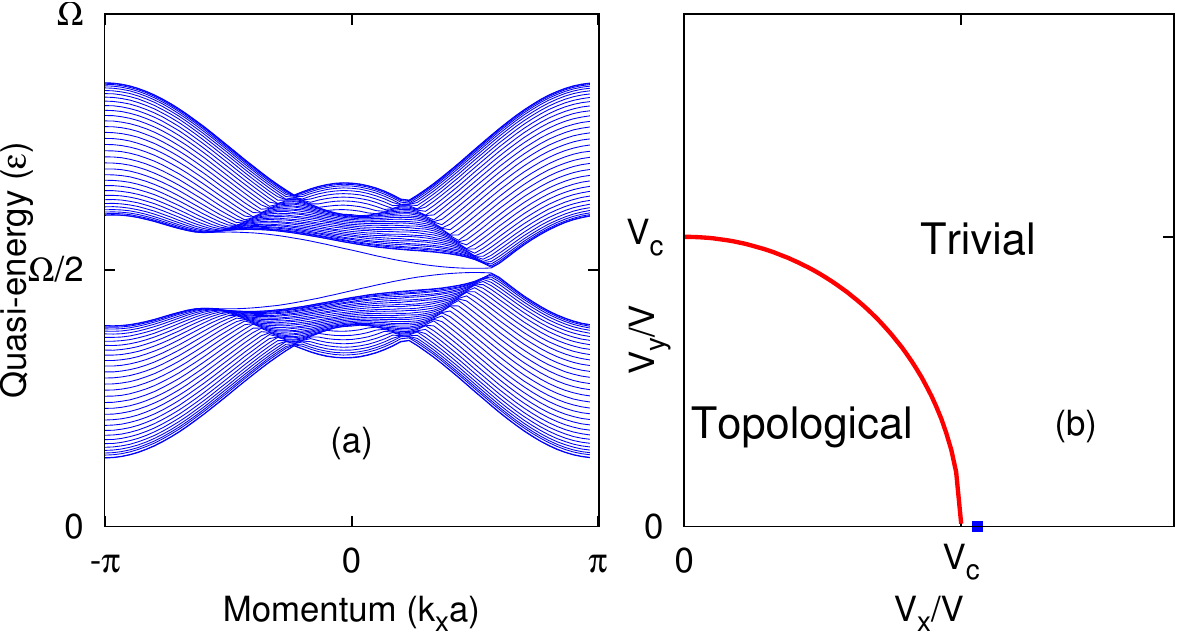}
  \caption{(a) Bandstructure of the driven system for $V_z/M=2$, $V_x/V_z=0.15$ in the trivial phase. The other parameters of the Hamiltonian is same as Fig. \ref{fig:Floquetbandstructure}. (b) shows the typical phase diagram for the Floquet topological phase as a function of $V_x$ and $V_y$. For $V_{x,y}<V_c$, the phase is topological and otherwise trivial. The clean system phase for obtaining the FTAI phase is chosen such that $V_x>V_c$, with $V_y=0$. The initial point of the trivial driven system is schematically represented as the point (blue square) in this diagram.}
    \label{fig:cleanFTAIphasediag}
 \end{figure}
 \subsection{Review: TAI and FTAI.}\label{subsec:ReviewTAI}
    Disorder counter intuitively may induce topological phases in a clean trivial system and such phases have been labeled as Topological Anderson insulators (TAI). As an explicit example we consider a trivial semiconducting quantum well described by the BHZ model, as shown in Eq. (\ref{eq:H_rad}) (with $V=0$). In this case, the clean system is trivial, i.e. $M/2B<0$. In contrast with trivial Anderson insulators which localize for any finite disorder strength, these models with spin-orbit coupling, on-site disorder in the chemical potential gives rise to the possibility of a localization transition at a finite disorder strength. Before the onset of localization, disorder primarily renormalizes the parameters of the Hamiltonian modifying the single particle density of states. Notably, this renormalization of the mass, $M$, leads to an induced topological phase.

     At weak disorder strength, sufficiently smaller than the critical disorder for the localization transition, there must exist a region of extended states in the band-structure. To lowest order in the disorder potential, the density of states of the  disordered Hamiltonian is given by the self-consistent Born approximation of the disorder-averaged Green's functions. In this approximation, the self-energy correction to the average Green's function\cite{BeenakkerTAI2009},
    \begin{eqnarray}
     \Sigma(E_F)&=&\overline{U^2}\int d{\bf k} \frac{1}{E_F-H_0({\bf k})-\Sigma}\nonumber\\
       &=& \Sigma_I \mathds{1}+\Sigma_x\sigma_x+\Sigma_y\sigma_y+\Sigma_z\sigma_z.
    \end{eqnarray}
Here, $\overline{U^2}=\sigma_d^2$, is second moment of the disorder potential, and $E_F$ is the fermi-energy. Note that this renormalization is present even if the average value of the disorder potential vanishes, $\overline{U}=0$. For uniformly random disorder, $[-U_0/2,U_0/2]$, an analytical formula is obtained using the Born approximation\cite{BeenakkerTAI2009}. The renormalized parameters of the model [see eq (\ref{eq:H_rad})] are
    \begin{eqnarray}
     \tilde{A}&=& A, \ \ \tilde{B} =B, \ \ \tilde{E}_F=E_F,\nonumber\\
     \tilde{M}(E_F) &=& M+\frac{U_0^2}{48 \pi}\frac{1}{B}\ln\left|\frac{B\pi^4}{E_F^2-M^2}\right|,
    \end{eqnarray}
 where, in general, the renormalized parameters  are functions of the fermi-energy, $E_F$, since the self energy correction due to disorder is, in general, a function of all the variables, $\Sigma\equiv \Sigma(E_F,A,B,M)$. In the trivial phase, $M>0$, and $B<0$, the contribution  due to disorder is negative, and when $\tilde{M}<0$, the phase becomes topological. We note that this phase transition is independent of the localization transition at strong disorder.

  Disorder induced topological phases have been shown to exist in the periodically driven honeycomb lattice models\cite{TitumPRL2015}. The phase in this driven system was labeled as the Floquet Topological Anderson Insulator (FTAI). The FTAI phase is a unique topological phase because it requires the presence of both the drive and disorder to exist. 
   Circularly polarized light\cite{Oka2009,Kitagawa2011} opens gaps at the Dirac points of the honeycomb model, akin to the gap in the Haldane model for the anomalous quantum Hall effect\cite{Haldanemod1988}. Perturbations which break inversion symmetry compete with this topological gap, and if strong enough, may result in a topologically trivial bandstructure.  The effect of disorder is to renormalize the gaps at the Dirac nodes in such a manner which induces a topological phase.

  \subsection{Realizing FTAI in quantum wells.}
 We now discuss how the FTAI can be realized in the semiconductor quantum-wells. For simplicity, we consider the case of generalized Zeeman field, with the components, $V_{x,y,z}$ are constants.  We take the Hamiltonian in Eq. (\ref{eq:H_rad}) and restrict ourselves to the case with a single resonance, $\left\{M,\frac{W}{2}\right\}<\Omega<W$. We choose the form of the drive (see below for the specific choice of $V$) so that the quasi-energy spectrum is topologically trivial, and add on-site uniformly random on-site disorder. In the following, we show that at finite disorder strength a transition to a topological quasi-energy spectrum occurs. We refer to this phase as the Floquet topological Anderson insulator.  As we show below, the mechanism underlying  the disorder-induced transition into the Floquet topological phase is different from the one underlying the TAI phase discussed in Sec. \ref{subsec:ReviewTAI}.  
  
  In the discussion below, we consider the Hamiltonian given by Eq.~(\ref{eq:H0+V}), corresponding to only one  $2\times 2$ block of the full BHZ model. It is natural to ask what type of topological phase can be induced by disorder in the full BHZ model driven by a generalized Zeeman field.  The disorder-induced topological phase of this block exhibits a non-zero Chern number, $\mathcal{C}_u=1$. A similar analysis carries over to the time-reversed $2\times 2$ block, for which the induced topological phase will exhibit a Chern number $\mathcal{C}_l=-1$ . The Floquet spectrum of the full BHZ model, driven with a generalized Zeeman field, is time reversal symmetric~\cite{Lindner2011},  and from the above considerations, we conclude that it will exhibit a disorder induced time-reversal invariant topological phase exhibiting  a non-zero $\mathds{Z}_2$ invariant.
   \begin{figure}
  \includegraphics[width=\linewidth]{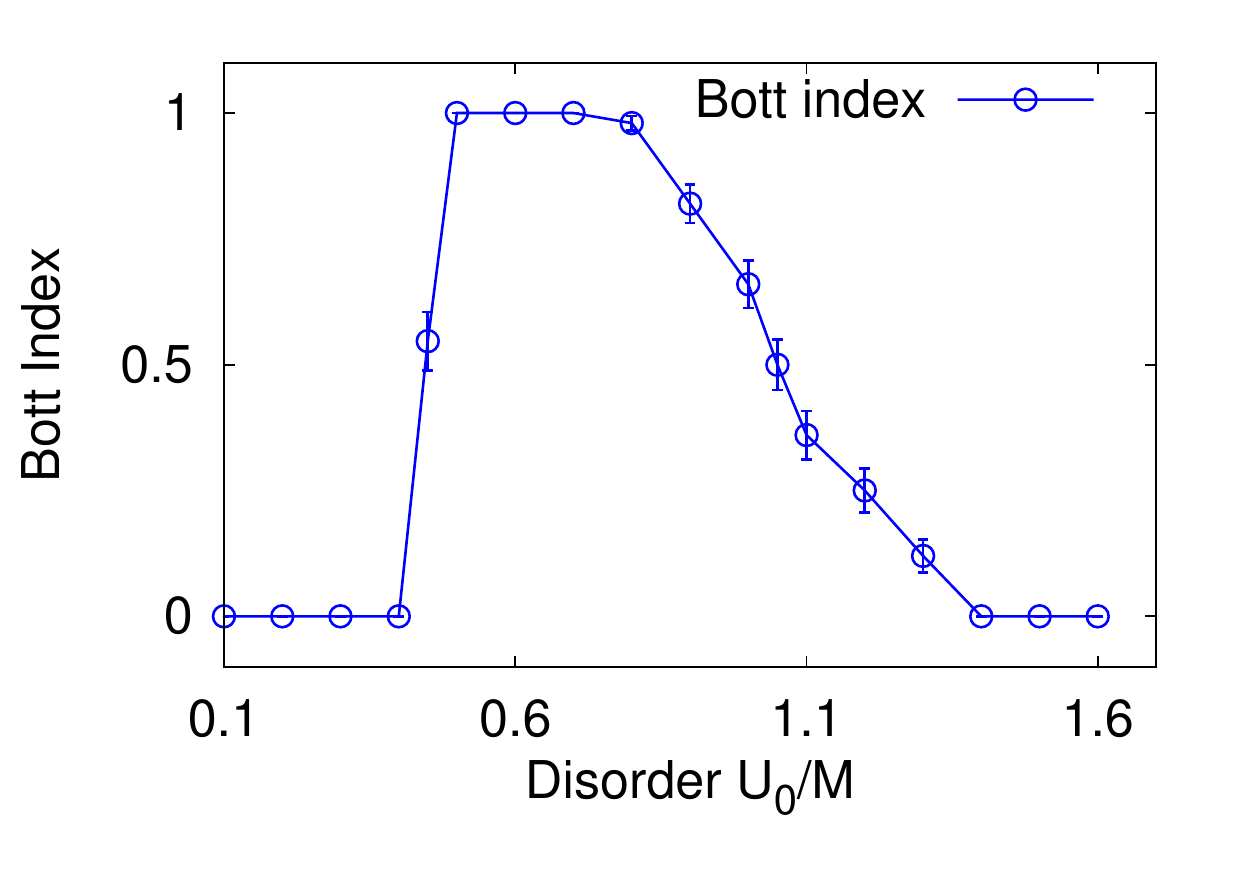}
  \caption{ Disorder-averaged Bott index, $\overline{C}_b(0,\Omega/2)$ as a function of disorder strength $U_0/M$. The clean system starts of as trivial, as evidenced by the band-structure (see Fig. \ref{fig:cleanFTAIphasediag}(a)) and the fact that the Bott index, $\overline{C}_b$ is 0 a small disorder. At finite disorder strength, the Hamiltonian acquires a non-zero average Bott index indicating the presence of a topological phase.The dimensions of the system considered are $L_x\times L_y=40\times 40$. }
    \label{fig:FATIbottindex}
 \end{figure}

 The phase of the driven model is always trivial for a large enough $V_x$ or $V_y$. This is in contrast to the topological phase induced when ${\bf V}=(0,0,V_z)$, as shown in Fig. (\ref{fig:Floquetbandstructure}). As discussed in section II, the quasi-energy gap at resonance ($\epsilon=\Omega/2$), is topological as long as the map defined by the vector $\hat{n}$ [see eq. (\ref{eq:nhat})] has a non-trivial Chern number. This condition is satisfied for,
 \begin{equation}
  \frac{V_{x,y}}{V}< V_c
 \end{equation}
where $V_c=|\sqrt{V_x^2+V_y^2}|$ is such that $\exists$ ${ \bf \tilde{ k}}$ in the Brilluoin zone, for which $|{\bf V}_\perp ({ \bf \tilde{ k}})|=0$. Intuitively, $|{\bf V_\perp}|$ [defined in Eq.(\ref{eq:Vperp})] is the magnitude of the gap that opens at resonance, $\epsilon=\Omega/2$, and therefore, at the critical condition, $V_{x}$ or $V_y=V_c$, this gap closes.  The schematic phase diagram for the topological phase in the presence of a radiation field, ${\bf V}$, as a function of $V_x$ and $V_y$ is shown in Fig. \ref{fig:cleanFTAIphasediag} (b).

Now consider the case where we choose the radiation potential, $V=(V_x,0,V_z)$, such that it lies in the trivial phase.For example, Fig. \ref{fig:cleanFTAIphasediag} (a) shows the band-structure with $V_x/V_z=0.15$ and $V_y=0$, in a cylindrical geometry. The bulk bands are gapped with no edge states crossing the gap, and this clearly indicates a trivial phase. 
Disorder renormalizes the parameters of the Floquet Hamiltonian. The gaps at the resonance are renormalized in a way to induce a topological phase. In the following section we show numerically, that a topological phase is obtained for sufficiently strong disorder, due to the renormalization of the radiation potential.  
 We note that it is essential to choose the parameters close to the phase transition. This is because, for a large enough, $V_x\gg V_c$, the bands localize before a topological phase is induced.  We also note that at the critical disorder strength at which the FTAI is induced, the static part of the model remains trivial. At this disorder strength, the renormalized mass parameter $\tilde{M}$ does not change sign. Therefore, in contrast to the TAI phase, the FTAI phase is not induced due to renormalization of the mass parameter $M$.

 \begin{figure}
  \includegraphics[width=\linewidth]{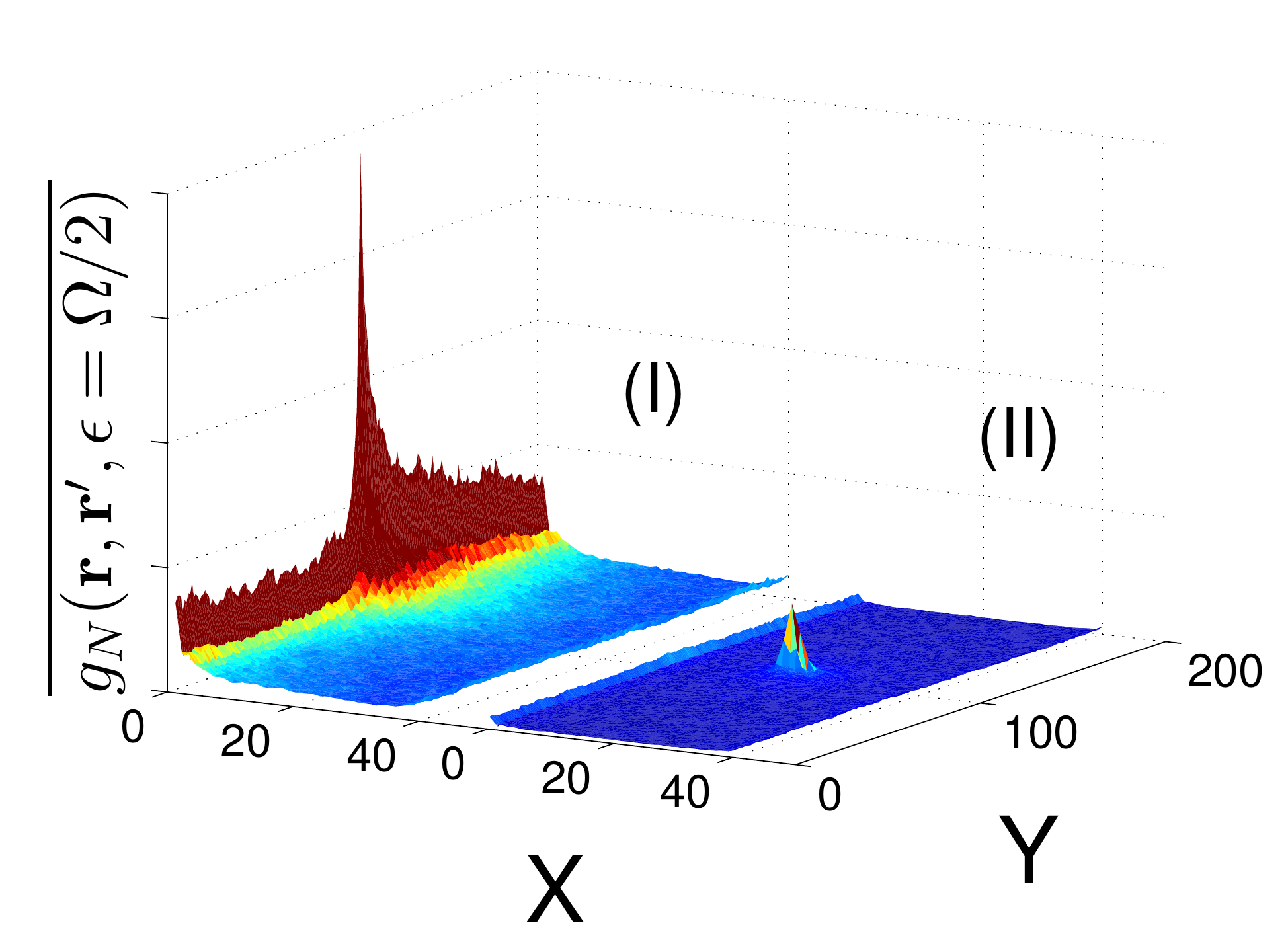}
  \caption{The time-evolution of a $\delta$-function wavepacket for disorder strength $U_0/M=0.7$. Figures (I) and (II) show $\overline{g_{N}(\epsilon=\Omega/2,{\bf r},{\bf r}')}$ for different choice of initial position ${\bf r}$. (I) shows the presence of an extended edge mode at quasienergy, $\epsilon=\Omega/2$. (II) on the other hand, shows that choosing the starting wave-packet in the bulk, continues to remain localized in the bulk. All simulations were carried out on a lattice of size $L_x\times L_y=40\times 200$ and for a total number of time periods, $N=5000$.}
  \label{fig:FATItimeevo}
 \end{figure}

 \subsection{Numerical Results}
We numerically examine the Bott index $\overline{C}_b(0,\Omega/2)$ as a function of disorder. The parameters of the radiation potential is chosen to be in the trivial phase, with $V_x/V_z=0.15$ and $V_y/V_z=0$, and $V_z/M=2$.  The 'particle-hole' symmetry around the resonance, $\epsilon=\Omega/2$, ensures that the trivial-to-topological transition must occur at the resonance quasi-energy. Fig. \ref{fig:FATIbottindex} shows the Bott index, $\overline{C}_b(0,\Omega/2)$ as a function of the disorder strength. At small disorder, the index is $0$ indicating a trivial phase. At intermediate disorder strength, the index $\overline{C}_b\sim 1$ implying that a topological phase is induced at this disorder strength. Strong disorder leads to localization of all Floquet eigenstates which is again topologically trivial indicated by $\overline{C}_b\rightarrow 0$. The topological nature of the induced phase may also  be detected from the emergence of non-trivial edge states. In the FTAI phase, this can be verified by an exact time-evolution of wave-packets on the edge as discussed in section III A. In Fig. \ref{fig:FATItimeevo}, we obtain the average transmission probability at quasi-energy, $\epsilon=\Omega/2$, for disorder strength $U_0/M=0.7$. For the clean system, this quasi-energy corresponds to the trivial-gap in the spectrum and has no bulk or edge states. In the disordered system, Case I in Fig \ref{fig:FATItimeevo} shows that there exists extended states on the edge, which was obtained when the initial wave-packet was chosen on the edge. On the other hand, case II shows that choosing the initial wavepacket in the bulk keeps the state localized around the initial state. This confirms that the state observed is indeed a topological state. 
\section{Realizing the FTAI using elliptically polarized light}
  In this section we outline a  proposal for realizing the Floquet topological Anderson Insulator phase by irradiating  semiconductor quantum wells using elliptically polarized light. In the preceding section, we introduced the FTAI phase using an idealized model of a generalized Zeeman field. We now show that by driving the system with elliptically polarized light, it is possible to induce a topological phase at finite disorder strength. Experimentally, it is easier to tune the polarization of light than controlling the impurity concentration. We show that it is possible to detect the presence of the FTAI phase by tuning the polarization of light illuminating a disordered sample.

Ref.~\onlinecite{Lindner2011} showed that semiconductor quantum wells subjected to circularly polarized light exhibit a Floquet spectrum akin to a Chern insulator, with a Chern number, $\mathcal{C}=2$. In the following, we focus on the Floquet spectrum of a single $2\times 2$ block of the BHZ model in the presence of circularly polarized light. On changing the ellipticity of the incident light, it is possible to tune the Floquet spectrum of this driven model from a topological to a trivial phase (See Appendix C for details). Since we are irradiating with elliptically polarized light, this model breaks time-reversal symmetry explicitly. This provides an ideal starting point to realize an analogous FTAI phase with broken time-reversal symmetry. By choosing a polarization such that the incident light induces a trivial phase, the FTAI phase can be obtained by adding disorder. As discussed in Section V, the disorder renormalizes the various components of the drive,. In this case, disorder renormalizes the ellipticity of the radiation, thereby inducing a topological phase at finite disorder strength which is the FTAI phase.

  An elliptically polarized light is introduced into the quantum well model, $H_0$ [see Eq. (\ref{eq:H_rad})] through a time-varying gauge field, $|{\bf A}|=\left(A_x \sin(\Omega t),A_y \cos(\Omega t)\right)$. The magnitude of $A_x$ and $A_y$ determine the ellipticity of the light. In the following, we always choose the magnitude of $|{\bf A}|=\sqrt{A_x^2+A_y^2}=1$, and the ellipticity is determined by $\theta=\arctan(A_y/A_x)$. In the perturbative regime, the Hamiltonian can be written in a generalized form analogous to that defined in Eq. (\ref{eq:H_rad}). In fact, it can  be shown that the unit vector, $\hat{n}$ defined in Eq. (\ref{eq:nhat}) winds around the sphere twice resulting in a topological phase with Chern number, $\mathcal{C}=2$. The details of the derivation of the quasi-energy gap and winding for this model are provided in the Appendix C.

  \begin{figure}
  \includegraphics[width=\linewidth]{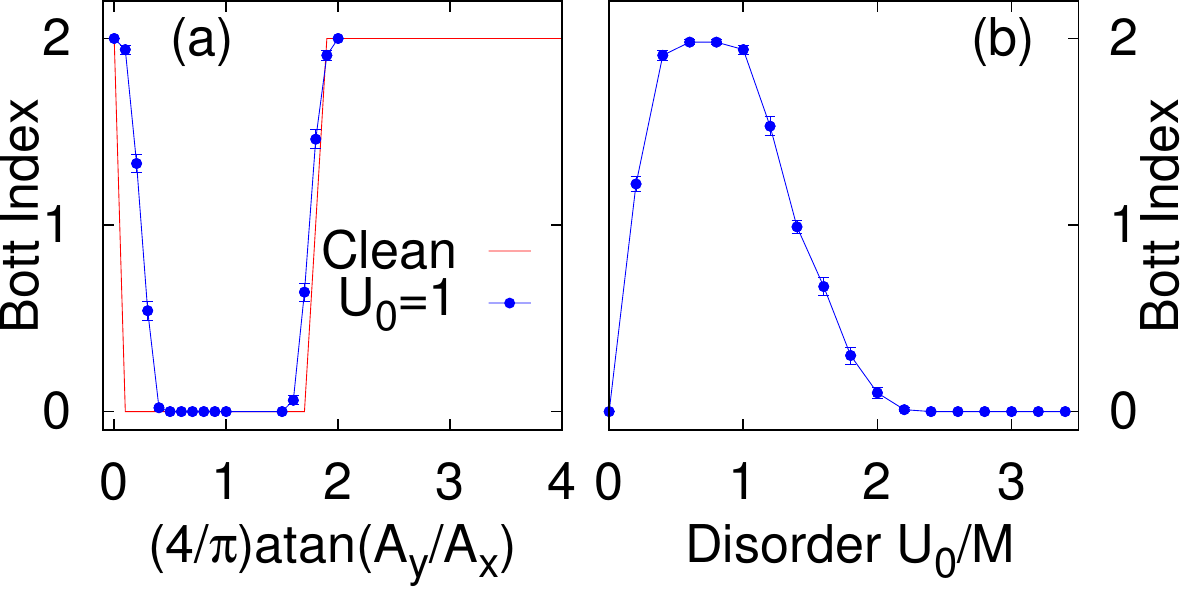}
  \caption{(a) Comparison of the Bott index, $C_b(0,\Omega/2)$ for as a function of the polarization angle, $\theta$ for the clean sample and disorder strength, $U_0/M=1$. Clearly, the region where the system is topological, $\overline{C}_b(0,\Omega/2)=2$, is larger in case of the disordered system, indicating the presence of a disorder-induced topological phase. (b) Bott index as a function of disorder strength when the initial polarization is $\theta =0.1\frac{\pi}{4}$. A topological phase, with $\overline{C}_b(0,\Omega/2)=2$, is induced as a function of disorder. The simulations were run for system sizes, $L_x \times L_y=20\times 20$.}
  \label{fig:circularlypolarizedFATI}
  \end{figure}
  We numerically compute the Bott index from the Floquet spectrum of this model. The undriven system is chosen with the following parameters, $A/M=1$, $B/M=-0.2$, and the driving frequency is $\Omega/M=3.1$. The time-dependent gauge field is also chosen such that, $|A|/M=1$ with varying $\theta$.  The topological nature of the Floquet bands is determined by the angle $\theta$. Fig.\ref{fig:circularlypolarizedFATI} (a) shows the  Bott index as a function of $\theta$. The Chern number of the band of the clean system is shown in red. Clearly, the clean system is trivial for $\theta=0.1\pi/4$. We choose this angle as the starting point for obtaining the FTAI. As shown in Fig \ref{fig:circularlypolarizedFATI} (b), at finite disorder, for this polarization angle, $\theta=0.1\pi/4$, a topological phase is induced, which we identify as the FTAI. Fig \ref{fig:circularlypolarizedFATI} (a) also shows the dependence of the disorder averaged Bott index at a disorder strength, $U_0/M=1$ on the angle $\theta$. A consequence of a disorder induced topological phase is that a larger fraction of the polarization angles, $\theta$, are topological. The existence of this phase may be detected experimentally by comparing a disordered and clean sample by tuning the polarization of incident light.

   So far, we have discussed a disorder induced Chern insulator phase in the Floquet spectrum of a single $2\times 2$ upper-block of the full BHZ model. The non-driven Hamiltonian of this model, corresponding to $H_0({\bf k})$ in Eq. (\ref{eq:H_rad}), has no time-reversal symmetry.  One possible way to experimentally observe the disorder induced topological phase is to isolate the bands of each block in the full BHZ model using a Zeeman coupling to a constant magnetic field. We now discuss the behavior of the full $4\times 4$ BHZ model in the presence of elliptically polarized light and disorder, which is more subtle. Let us define the Chern numbers for the upper and lower blocks of the BHZ model as $\mathcal{C}_u $ and $\mathcal{C}_l$ respectively. The lower block is the time-reversed counterpart of $H_0$, $H_0^*(-{\bf k})$. It has a Chern number, $\mathcal{C}_l=-2$ when the polarization angle, $0<\theta<\frac{\pi}{2}$. As a result, when we consider the clean system, at $\theta=0.1\frac{\pi}{4}$, the bands of the full BHZ Hamiltonian are topological with a Chern number $\mathcal{C}=\mathcal{C}_l+\mathcal{C}_u=-2$. Now, when a FTAI phase with $\mathcal{C}_u=2$ is induced at finite disorder in the upper block, for $\theta=0.1\pi/4$, the lower block is still robust, and has a Chern number, $\mathcal{C}_l=-2=-\mathcal{C}_u$. This means that counterpropagating edge modes exist at quasi-energies around $\Omega/2$. There is no symmetry to prevent scattering between these counterpropagating states. So, an infinitesimal coupling between the two sets of counterpropagating edge modes is going to gap them out. Therefore, the system with the upper-band FTAI phase is actually a trivial insulator. The induced topological phase in the upper block manifests as a topological to trivial transition for the full BHZ model.
\section{Conclusions}

In this paper we studied the effects of quenched disorder on the quasi-energy spectrum of Floquet topological insulators realized within the semiconductor quantum well model. In this model, the driving frequency is resonant with a transition between the valence and conduction band. In the first part of the paper, we examined the robustness of the topological properties of the Flouqet spectrum when quenched, on site disorder is added.  We showed that these topological properties remain robust as long as there exists a mobility gap in the quasi-energy spectrum, centered around the resonance (which in our conventions occurs at quasi-energy $\Omega/2$). This behaviour can be understood by recalling that quasi-energy bands with non-zero Chern numbers exhibit a delocalized state. Consequently, the topological phase is robust so long as there is a finite mobility gap between these two delocalized states. Such a localization transition is analogous to that occurring in quantum hall systems and Chern insulators.

In the second part of the paper we showed that periodically driven semiconductor quantum wells can host disorder induced topological phases. Such Floquet topological Anderson insulators (FTAIs)  are topological only in the presence of both periodic driving and sufficiently strong disorder. Absent the disorder, the driven system is topologically trivial:  the parameters of the drive are chosen to induce a trivial quasi-energy gap. When sufficiently strong disorder is introduced, it renormalizes the parameters of the drive  and induces a transition to a topological phase.
 
\begin{acknowledgments}
P.T., N.H.L., and G.R. acknowledge support from the U.S.-Israel Bi-National Science Foundation (BSF). G.R. and P.T. are grateful for support from NSF through DMR-1410435. G.R. would like to acknowledge Aspen Center for Physics for hospitality under NSF grant 1066293. N.H.L. acknowledges support from I-Core, the Israeli excellence center Circle of Light, from the People Programme (Marie Curie Actions) of the European Union’s Seventh Framework Programme (FP7/2007-2013) under REA Grant Agreement No. 631696, and from the European Research Council (ERC) under the European Union Horizon 2020 research and innovation programme, under Grant Agreement No. 639172. P.T. was also was supported by the AFOSR, ARO MURI, ARL CDQI, NSF QIS, ARO, and NSF PFC at JQI.
\end{acknowledgments}

\section*{Appendix}
\appendix
\section{Born Approximation}
  The correction to the density of states of the quasi-energy band-structure in the presence of dilute disorder is obtained perturbatively in the Born approximation. We generalize the method used for time-independent systems to the driven Hamiltonian by obtaining the self energy correction to the Floquet Green's function as a result of disorder. In the following, we first describe the formalism to obtain Floquet Green's functions as an expansion in the the radiation potential $|V|$. Next, we calculate the self energy correction to the Green's function in the presence of disorder.
 \subsection{Floquet Green's function}
\begin{figure}
 \includegraphics[width=\linewidth]{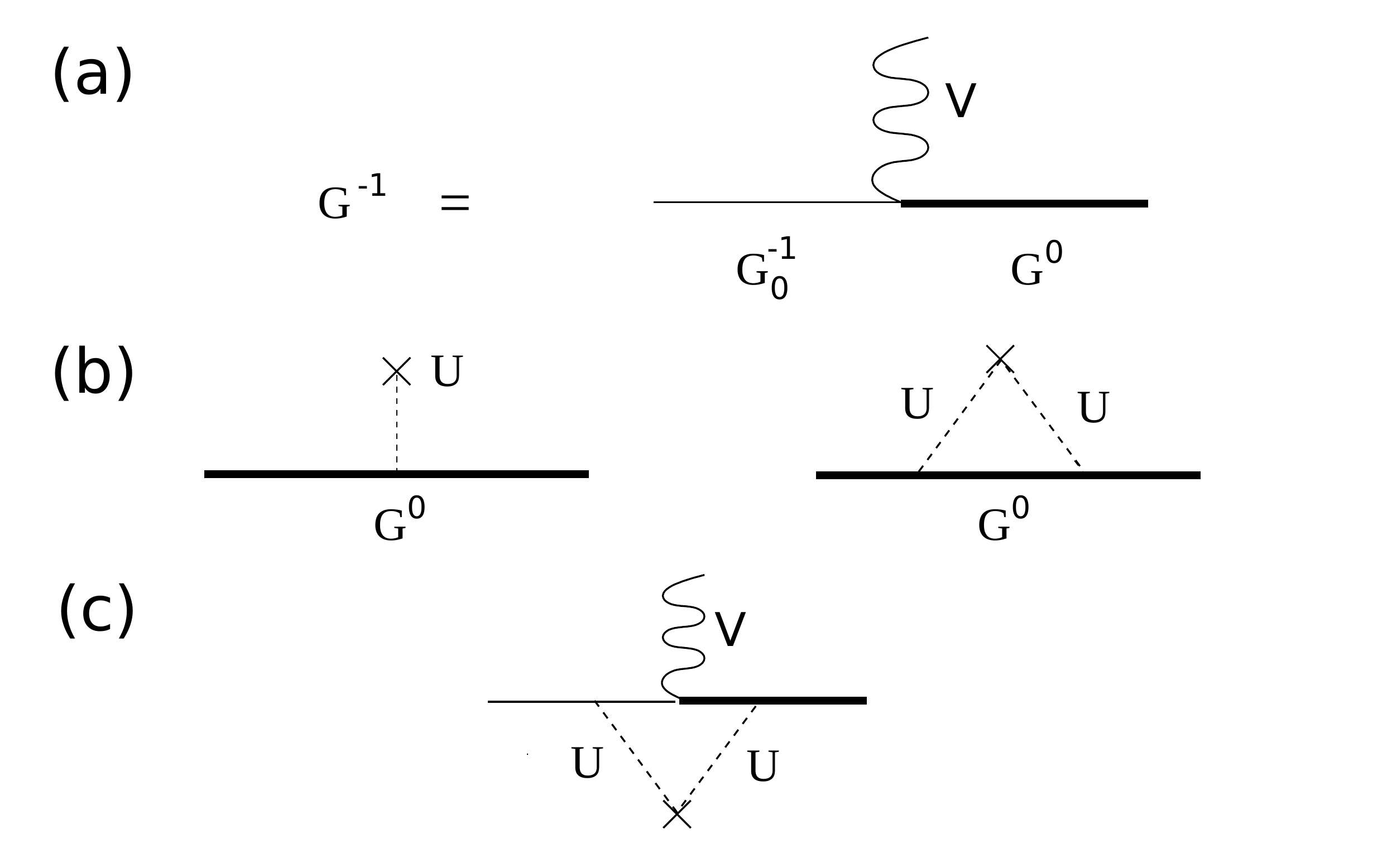}
  \caption{This figure illustrates the diagramatic representation of the Floquet Green function and the renormalization due to disorder. The thin line represents the bare propagator defined in Eq. (\ref{eq:barepropagator}). The thick line represents the propagator in the presence of the drive, $G^0$, defined in Eq. (\ref{eq:G^0}). (a) The $n=-1$ component, $G^{-1}$ as defined in Eq. (\ref{eq:G-1}). (b) The renormalization due to disorder of $G^0$. This corresponds to renormalization of the bare parameters of the time-independent part of the Hamiltonian. (c) The self energy correction due to disorder to the radiation potential.  The change of the density of states near the Floquet gap at resonance is due to the renormalization of the radiation potential.  }\label{fig:feyndiag}
 \end{figure}

 The Green's function is defined for the clean Floquet Hamiltonian, $H_r^F$ (see Eq. (\ref{eq:FloquetMatrix})),
 \begin{widetext}
\begin{equation}
 G^F_r(\epsilon, {\bf k})=\frac{1}{\epsilon-H_r^F({\bf k})}
\end{equation}
This is formally an infinite dimensional matrix. In the extended zone scheme, the Floquet Green function is written as,
  \begin{equation}
   G^F_r=\left( \begin{array}{cccccc}
               \ddots&\ddots &\vdots &&& \\
                G^{-1}(\epsilon+\Omega,{\bf k})&G^{0}(\epsilon+\Omega,{\bf k}) &G^{1}(\epsilon+\Omega,{\bf k}) &\udots&  \\
                  & G^{-1}(\epsilon,{\bf k})&G^{0}(\epsilon,{\bf k}) &G^{1}(\epsilon,{\bf k})& \\
                  & \udots & G^{-1}(\epsilon-\Omega,{\bf k}) &G^{0}(\epsilon-\Omega,{\bf k}) &G^{1}(\epsilon-\Omega,{\bf k}) \\
                  && \vdots   & \ddots\\
              \end{array} \right), \label{eq:FloquetGreenMatrix}
  \end{equation}
  \end{widetext}
  where the different components $G^n(\epsilon,\textbf{k})$ is obtained in the perturbative regime with the radiation potential, $V={\bf V}\cdot{\boldsymbol \sigma}/2$ with $|V|/\Omega \ll 1$. In the following, we write the components of $G^F_r$, order by order in the radiation potential.  Let us start by defining the matrix elements of the bare Floquet Green's functions in terms of the the time-independent Hamiltonian, $H_0$ as,
\begin{eqnarray}
 [G^F_0(\epsilon,{\bf k})]_{mn}&=&\frac{\delta_{mn}}{\epsilon-H_0-n\Omega}\\
 &\equiv& \delta_{mn} G_0^{-n}(\epsilon,\textbf{k}),\label{eq:barepropagator}
\end{eqnarray}
where ${G}^F_0$ is exactly the Floquet Green's function to zeroth order (setting $V=0$). The different components of the Floquet Green function can be written down as a continued fraction \cite{Martinez}. Let us define two operators,
\begin{eqnarray}
\mathcal{F}_{+}^n\left(\epsilon\right)&=&\frac{1}{\left(G_{0}^n\left(\epsilon\right)\right)^{-1}-V\frac{1}{\left(G_{0}^{n+1}\left(\epsilon\right)\right)^{-1}-\vdots}V}\text{ for }n>0\ \ \ \ \ \ \label{eq:Fplus}\\
\mathcal{F}_{-}^n\left(\epsilon\right)&=&\frac{1}{\left(G_{0}^n\left(\epsilon\right)\right)^{-1}-V\frac{1}{\left(G_{0}^{n-1}\left(\epsilon\right)\right)^{-1}-\vdots}V}\text{ for }n<0,\ \ \ \ \ \label{eq:Fminus}\
\end{eqnarray}
where, we omitted the explicit functional dependence on $\textbf{k}$. Now, we can write down all the components of the Green functions,
\begin{eqnarray}
G^0\left(\epsilon\right)&=&\frac{1}{\epsilon-H_{0}-V_{\text{eff}}}\\
V_{\text{eff}}&=&V\mathcal{F}^{-1}_{-}\left(\epsilon\right)V+V\mathcal{F}^{1}_{+}\left(\epsilon\right)V\\
G^n\left(\epsilon\right)&=&\begin{cases}
\mathcal{F}^n_{+}\left(\epsilon\right)V\cdots\mathcal{F}^1_{+}\left(\epsilon\right)VG^0\left(\epsilon\right),\text{ for }n>0\\
\mathcal{F}^{n}_{-}\left(\epsilon\right)V\cdots\mathcal{F}^{-1}_{-}\left(\epsilon\right)VG^0\left(\epsilon\right),\text{ for }n<0\end{cases}
\end{eqnarray}
As shown in Section III, we  consider parameter ranges which induces only one resonance in the bandstructure, and therefore, are concerned with the effect of disorder at quasi-energies near the resonance. A weak resonant drive only mixes adjacent Floquet bands appreciably, and therefore it is sufficient to consider only the $n=0$ and $n=1$ components of the Green function, $G^n(\epsilon,\textbf{k})$.  In order to make analytical progress, we truncate the continued fraction expressions for $\mathcal{F}_\plus$ and $\mathcal{F}_-$ given in Eq. (\ref{eq:Fplus}) and (\ref{eq:Fminus}), at the zeroth order to obtain,
\begin{eqnarray}
\mathcal{F}_{+}^{1}\left(\epsilon\right)&=&G_{0}^1\left(\epsilon\right),\label{eq:F1plus}\\
\mathcal{F}_{-}^{-1}\left(\epsilon\right)&=&G_{0}^{-1}\left(\epsilon\right), \label{eq:F1minus}\\
 G^0\left(\epsilon\right)&=&\frac{1}{\epsilon-H_{0}-V_{\text{eff}}},\label{eq:G^0}\\
G^{-1}\left(\epsilon\right)&=&
\mathcal{F}^{-1}_{-}\left(\epsilon\right)VG^0\left(\epsilon\right).\label{eq:G-1}
\end{eqnarray}
The different components of the Floquet Green function is represented using Feynman diagrams shown in Fig. \ref{fig:feyndiag}. We note, that in this truncation scheme, we are dropping all terms involving $G_0^n$, $|n|>1$. This is valid when only processes involving single photon are relevant. In the BHZ model we have considered, we assume that the drive induces only one resonance in the band-structure.
 \subsection{Floquet Born approximation}
  The disorder potential can now be treated perturbatively using the Born approximation. The disorder-averaged Floquet Green's function is,
 \begin{equation}
\overline{ G^F_r({\bf k},\epsilon)}=\frac{1}{\epsilon-H^F_r-\Sigma^F},
 \end{equation}
 where, $\Sigma^F$ is the self-energy correction due to disorder and $\overline{(\cdots)}$ indicates disorder averaging. In the Born approximation, the self energy is,
\begin{equation}
\Sigma^F=\overline{U^2}\int_{FBZ}d{\bf k}\ G^F_r(\epsilon,\textbf{k}).\label{eq:vanillaborn}
\end{equation}
The contribution of the self energy can be split into different Floquet blocks, similar to the Floquet Green's function as shown in Eq. (\ref{eq:FloquetGreenMatrix}),
\begin{equation}
 \Sigma^F=\left( \begin{array}{cccc}
               \ddots &\ddots&&   \\
               \ddots&\Sigma^F_0& \Sigma^F_V& \\
               &\Sigma^{F\dagger}_V&\Sigma^F_1&\ddots\\
               &&\ddots &\ddots   \\
              \end{array} \right),\label{Sigma-mat}
\end{equation}
where $\Sigma^F$ renormalizes the Hamiltonian, $H^F_r$. The terms $\Sigma^F_{n}$ renormalizes the time-independent Hamiltonian, $H_0$, and $\Sigma^F_V$ renormalizes $V$. Therefore, the effect of disorder on the single-particle Green's function is two fold. First, it renormalizes the parameters of the bare, time-independent Hamiltonian $H_0$, $M\rightarrow \tilde{M}$, and $B\rightarrow \tilde{B}$. This is due to the diagonal terms in the self energy, $\Sigma_{n}^F$. Note that, at lowest order, this correction is identical to that from disorder-averaging the Green's function of the time-independent Hamiltonian [see Fig. \ref{fig:feyndiag} (d)]. This renormalization of the parameters is not important to the modification of the quasi-energy density of states near the resonance ($\epsilon\sim \Omega/2$). We assume that this effect can be neglected. The second effect is to renormalize the radiation potential, ${\bf V}\rightarrow \tilde{\textbf{V}}$. Since the magnitude of the Floquet band-gap, $V_g$, of the Hamiltonian is proportional to $|{\bf V}|$, we will primarily be interested in this particular renormalization.

 The diagram shown in Fig. \ref{fig:feyndiag} (c) is the leading order contribution to the self energy correction to the radiation potential. $V$. Explicitly, in the Born approximation, using Eqs. (\ref{eq:G-1}) and (\ref{eq:vanillaborn}) we obtain,
\begin{eqnarray}
\Sigma_V&=&\overline{ U^2}\int_{FBZ}d{\bf k}\ \ G^{-1}(\epsilon,\textbf{k}) \\
&=&\overline{ U^2}\int_{FBZ}d{\bf k}\ \ \mathcal{F}^{-1}_{-}\left(\epsilon\right)VG^0\left(\epsilon\right)\label{eq:born-approx-integral}\\
&=&\Sigma_V^I \mathds{1}+\sum_{i=x,y,z} \Sigma_V^i \sigma^i \label{eq:self-energy-expansion}
\end{eqnarray}
where, $\overline{ U^2}=\sigma_d^2=\frac{U_0^2}{12}$, is the variance of the uniformly random distribution. In Eq. (\ref{eq:born-approx-integral}), we expand the $2\times 2$ matrix, $\Sigma_V$, in Pauli matrices. The components in the expansion, $\Sigma_V^{x,y,z}$ determines  the renormalization of the different components of $\textbf{V}\equiv (V_x,V_y,V_z)$.

Let us now use the form of the Hamiltonian, $H_0$, defined in Eq. (\ref{eq:H_rad}). We have the following expressions for the bare Green's functions,
\begin{eqnarray}
 G_0^0(\epsilon,{\bf k})&=&\frac{\epsilon +{\bf d}({\bf k})\cdot {\boldsymbol \sigma}}{\epsilon^2-d^2}\label{eq:G_00},\\
 G_0^{-1}(\epsilon,{\bf k})&=&\frac{(\epsilon-\Omega) +{\bf d}({\bf k})\cdot {\boldsymbol \sigma}}{(\epsilon-\Omega)^2-d^2},\label{eq:G_01}
\end{eqnarray}
where, ${\bf d}({\bf k})\equiv(d_x,d_y,d_z)=(A\sin k_x,A\sin k_y,M-2B(2-\cos k_x-\cos k_y)$ and $d^2=d_x^2+d_y^2+d_z^2$.

\begin{figure}
 \includegraphics[width=\linewidth]{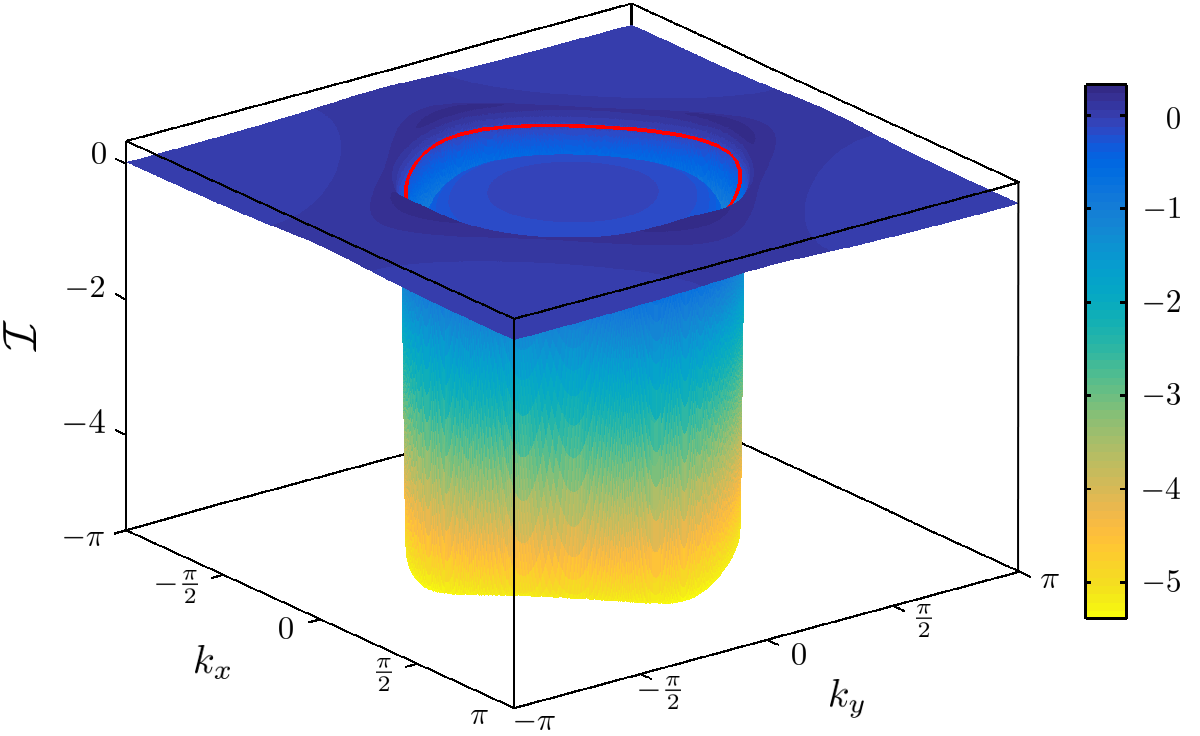}
 \caption{We plot the $\sigma_z$ component of the integrand, $\mathcal{I}$, defined in equation Eq. (\ref{eq:Sigma_z}) as a function of $k_x$ and $k_y$. The parameters are chosen with $V_z/M=0.1$, $A/M=0.2$, $B/M=-0.2$, $\Omega/M=3$. Clearly, the integrand is sharply peaked at the resonance circle, $d_r=d-\frac{\Omega}{2}=0$ (shown in red). The large negative value of the integrand results in the negative sign for $\Sigma^z_V$. We confirm this by a numerical integration the $\sigma_z$ component over the entire Brilluoin zone. }
 \label{fig:born_approx_integrand}
\end{figure}

Now, let us compute the correction due to the disorder potential, when the radiation field is along the $z$ direction, ${\bf V}=(0,0,2V_z)$. We note that the static Hamiltonian, $H_0$ (see Eq. (\ref{eq:H0+V})) is particle-hole symmetric. Consequently, the time-evolution operator for a single time-period and the Floquet Hamiltonian inherit this symmetry. As a result, the density of states is symmetric around the quasi-energy, $\epsilon=\Omega/2$, as is obvious from the spectrum shown in Fig. \ref{fig:Floquetbandstructure} (b). Since the disorder potential has a zero mean, $\overline{U}=0$, the disorder-averaged density of states must remain symmetric around $\epsilon=\Omega/2$. Therefore, the quasi-energy gap must close at $\epsilon=\Omega/2$. This is also confirmed numerically in Fig. \ref{fig:2.5} (a). Now, we can write down the expression for the self energy at the quasi-energy $\epsilon=\Omega/2$,
\begin{widetext}
\begin{eqnarray}
 \Sigma_V^z\left(\epsilon=\frac{\Omega}{2}\right)&=&\frac{U_0^2}{48\pi^2}\int d^2{\bf k}\underbrace{\frac{4V_{z}\left(+9\Omega^{4}+8\Omega^{2}\left(4d^{2}-9d_{z}^{2}\right)-16d^{2}\left(d^{2}-2d_{z}^{2}\right)+V_{z}^{2}\left(-32d^{2}+24\Omega^{2}\right)\right)}{Q}}_{\mathcal{I}}\sigma_z\label{eq:Sigma_z}\\
 \Sigma_V^I\left(\epsilon=\frac{\Omega}{2}\right)&=&\frac{U_0^2}{48\pi^2}\int d^2{\bf k}\frac{128\Omega d_{z}V_{z}^{3}}{Q}I,  \label{eq:Sigma_I}\\
 \text{with,}\ \ Q&=& 16\left(4d^{2}-9\Omega^{2}\right)d_{r}^{4}+16V_{z}^{2}\left(16d_{r}^{4}-16\Omega^{2}d_{r}^{2}-32d_{r}^{2}d_{z}^{2}-32\Omega^{2}\left(\frac{\Omega^{2}}{4}-d_{z}^{2}\right)\right)-256V_{z}^{4}d_{r}^{2}\\
 \text{and,}\ \ d_r^2&=&d^2-\Omega^2/4
\end{eqnarray}
\end{widetext}
where, we introduced $d_r$ to make notations compact. We note that at the resonance circle, $d_r=0$. Crucially, in this case when $V_x=V_y=0$ there is no contribution to $\Sigma^x_V$, $\Sigma^y_V$ since the integral vanishes by symmetry. This also means that disorder does not generate $x$ or $y$ components of the radiation if it is initially absent. We also note that the dominant contribution in the limit of weak radiation potential, $V_z/M\rightarrow 0$ is to $\Sigma_z$. In order to make progress with the integration in Eq. (\ref{eq:Sigma_z}), we note that the integrand, $\mathcal{I}$, is sharply peaked around the resonance circle, $d_r=0$. This is clearly seen in Fig. (\ref{fig:born_approx_integrand}).
Therefore, we compute the integration using a saddle point method by expanding the integrand around the maximum at the resonance circle. The integration in Eq. (\ref{eq:Sigma_z}) gives us,
\begin{equation}
\Sigma^{z}_V =-\overline{U^{2}}\frac{V_{z}}{2}\left[\frac{\frac{\Omega^{2}}{4}-d_{z}^{2}+\frac{\Omega^{2}}{4}V_{z}^{2}}{V_{z}^{2}\left(\frac{\Omega^{2}}{4}-d_{z}^2\right)}\right]\times\sqrt{\frac{\pi}{|\alpha|}},\label{eq:SigmaVsimplified}
\end{equation}
where $|\alpha|$ determines the width of the peak at the resonance circle and can be estimated from the second derivative of the integrand, $\alpha=\frac{\partial^2\mathcal{I}}{\partial k^2}{\big |}_{d=\frac{\Omega}{2}}$.  Clearly, at the resonance circle, $d_x^2+d_y^2+d_z^2=\Omega^2/4$, the sign of the correction is negative, $\Sigma_V^z<0$. This means the radiation potential which renormalizes to $\tilde{V}_z = V_z+\Sigma_V^z$, vanishes for large enough disorder. Since the quasi-energy gap is proportional to $|{\bf V}|$, this agrees with the fact that the gap closes as a function of disorder.  This expectation is verified by numerically performing the integration in Eq. (\ref{eq:Sigma_z}). The parameters chosen are $V_z/M=0.1$, $A/M=0.2$, $B/M=-0.2$, $\Omega/M=3$, for which we obtain, $\Sigma^z_V\approx-0.35\overline{U^{2}}$. Note that the calculation done in this section is valid only in the limit $V_z/M\rightarrow 0$. Several steps rely on this assumption, most notably the truncation of the continued fraction in Eq. (\ref{eq:F1plus}-\ref{eq:G-1}). The numerical simulations done in the main text is outside the range of this calculation. Nevertheless, we see effects qualitatively consistent with what we see from the perturbative analysis, namely, the quasi-energy gap is renormalized to zero as a function of disorder as seen in Fig. \ref{fig:2.5} (a). This calculation can be made more precise by obtaining $\Sigma_z$ in the self-consistent Born approximation, where all the parameters in Eq. (\ref{eq:Sigma_z}), are the disorder renormalized counterparts and not their bare values. A similar calculation can be done when the radiation potential is along the $x$ direction, $\textbf{V}=(2V_x,0,0)$. In this case, the disorder potential also renormalizes the potential, $V_x$.

\section{Numerical Methods}
In this appendix, we provide details of the numerical methods implemented to study the disordered periodically driven systems.
\subsection{Real-time evolution}
 This method is used to numerically obtain the disorder-averaged single-particle Green's function to high accuracy, which gives us the average transmission probability at a particular quasi-energy. The starting wave-packet is a $\delta$-function in real space, and we study the spreading of this wavepacket in real space as a function of disorder and quasi-energy. The bulk or edge properties are determined according to the choice of the initial position of the wavepacket and boundary conditions. While the bulk Green's function is obtained from the Hamiltonian in a torus geometry with periodic boundary conditions in both the $x$ and $y$ directions, the edge Green's function is obtained in a cylindrical geometry with open boundary conditions along $y$. In our simulations we start in a particular pseudospin, at position ${\bf x}$, $|{\bf x},\uparrow\rangle$, where $\sigma_z|\uparrow\rangle=|\uparrow\rangle$. The  Green's function is given by
 \begin{equation}
  G({\bf x},{\bf x}',t)=\langle{\bf x}', \uparrow|\exp(-i\int_0^t H(t') dt')|{\bf x},\uparrow \rangle,
 \end{equation}
 where $H(t)$ is the full Hamiltonian including the disorder potential and $\exp(-i\int_0^t H(t') dt')=U(t,0)$, is the time-evolution operator.

  The time-evolution operator is obtained numerically using a split operator decomposition. We take advantage of the facts that the Hamiltonian for the clean system, $H_0$ is a $2\times2$ matrix in the momentum space, and the disorder potential, $V_{\rm dis}$ is diagonal in real-space. We have for an infinitesimal time-step, $dt$, the time-evolution operator as,
  \begin{eqnarray}
   U(dt,0)&=&\exp\left(-iHdt\right)\nonumber \\
      &=& e^{-iH_0dt/2}e^{-iV_{\rm dis}dt}e^{-iH_0dt/2}+O(dt^3) \label{eq:U(dt)}
  \end{eqnarray}
Since the exponentiation of a diagonal or a small matrix is efficient, this method gives us an accurate way to obtain the exact time-evolution operator efficiently.

In our numerical simulations, we evolve the initial wavepacket for $N$ time-periods. At integral time-periods, the  time-evolution operator is identical to that of the Floquet Hamiltonian,
\begin{equation}
 U(NT,0)= \exp \left(\int_0^{NT}Hdt'\right)\equiv e^{-iH^FNT},
\end{equation}
where we note the fact that the Floquet Hamiltonian, $H^F$, is effectively 'time-independent'. Therefore, the time-evolution of a $\delta$-function wavepacket for $N$ time periods, gives us the Floquet Green's function in real space.
\begin{eqnarray}
 G({\bf x},{\bf x}',NT)&=&\langle {\bf x}' \uparrow|U(NT,0)|{\bf x}\uparrow\rangle \\
 G_{N}({\bf x},{\bf x}',\epsilon)&=&\int_0 ^{NT} dt\ \ G({\bf x},{\bf x}',NT)e^{-i\epsilon t}\nonumber \\
 &=&\langle {\bf x}' \uparrow|(\epsilon -H^F)^{-1}|{\bf x}\uparrow\rangle, \label{Beq:G^F}
\end{eqnarray}
where in Eq. (\ref{Beq:G^F}), we Fourier transformed in time to obtain the Floquet Green's function as a function of the quasi-energy. The subscript $N$, in Eq. (\ref{Beq:G^F}) refers to the total time of evolution, $T_{\rm tot}=NT$.

\begin{figure}
  \includegraphics[width=\linewidth]{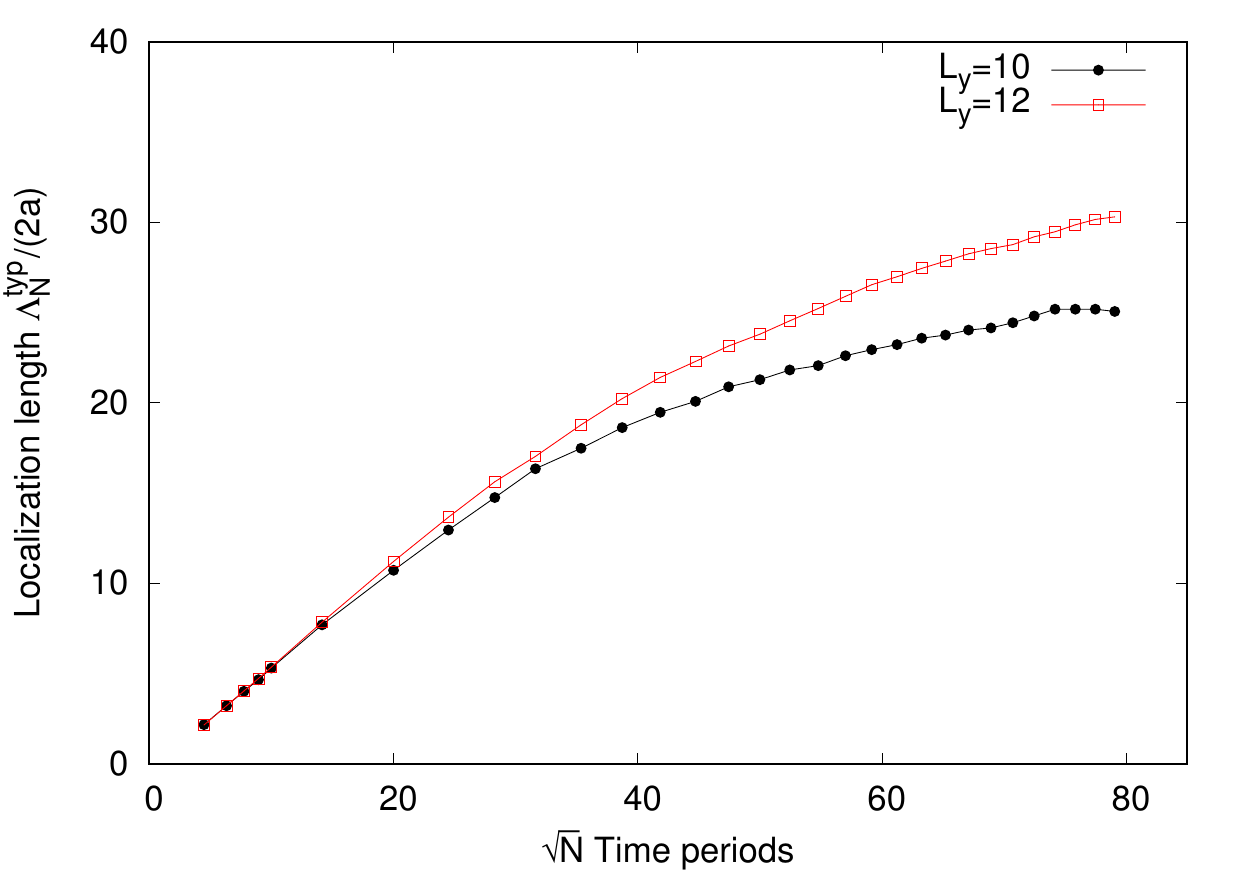}
  \caption{Typical localization length $\Lambda^\text{typ}_N/(2a)$ as a function of total time of evolution, $NT$, where $a=1$ is the lattice spacing.  (a) shows the typical bulk localization length at quasi-energy, $\epsilon=\Omega/2$, for $L_y=10$, and $12$. The disorder strength is $U_0/M=0.9$. For small times, the particle is diffusive, and at long times saturates to $\Lambda^\text{typ}_N=\Lambda_{\rm sat}$. $\Lambda_{\rm sat}$ scales with $L_y$ indicating that the state is extended.}
  \label{fig:timedependenceoflocalizationlength}
 \end{figure}

In order to study the effect of disorder, we must average the Green's function for a large number of disorder configurations.  To the study the properties of the system, and the transition to localization, we obtain the average transmission probability in a strip geometry. Initializing at ${\bf x}\equiv(x,y)$, we calculate the transmission probability to ${\bf x}'\equiv (x',y')$ as a function of disorder and quasi-energy,
\begin{eqnarray}
 \tilde{g}_{N}({\bf x},{\bf x}',\epsilon)&=&\overline{|G_N({\bf x},{\bf x}',\epsilon)|^2}, \label{eq:B15}\\
\end{eqnarray}
where $\overline{( \cdots)}$ denotes disorder averaging. We obtain the transmission probability along the $x$ direction, by summing over the $y$ direction,
\begin{equation}
 g_{N}({\bf x}, x',\epsilon)=\sum_{y'}\tilde{g}_{N}({\bf x},{\bf x}',\epsilon).
\end{equation}
 The properties of the $g_N$, are used to identify a localization transition in this system. $g_N$ is first obtained for a strip geometry of size $L_x\times L_y$, with $L_x \gg L_y$. We define a localization length along the $x$ direction,
\begin{equation}
\frac{1}{\Lambda_N(\epsilon)}= \frac{\sum_{x'}g_N({\bf x},x',\epsilon)^2}{\left(\sum_{x'}g_N({\bf x},x',\epsilon)\right)^2},
\end{equation}
 where $\Lambda_N\equiv \Lambda_{N,L_y}$ depends on the width of the strip $L_y$. Analogously, we can define the localization length for a typical disordered sample,
 \begin{equation}
  \frac{1}{\Lambda^\text{typ}_N(\epsilon)}= \frac{\sum_{x'}g^\text{typ}_N({\bf x},x',\epsilon)^2}{\left(\sum_{x'}g^\text{typ}_N({\bf x},x',\epsilon)\right)^2},
 \end{equation}
with $g_N^{\text{typ}}=\exp\left[\overline{\ln\left(\sum_y'G_N({\bf x},{\bf x}',\epsilon)\right)}\right]$, and $\overline{(\cdots)}$ denotes disorder averaging. In the following, we discuss the dependence of the typical localization length, $\Lambda^\text{typ}_N$  as a function of the width of the strip, $L_y$ and time of evolution $N$. The finite-size scaling of $\Lambda^\text{typ}_{N,L_y}(\epsilon)$ determines whether the particular eigenstate at quasi-energy $\epsilon$ is localized or extended. Extended states must scale with the system size with $\Lambda^{\text{typ}}_{L_y}/L_y\rightarrow \Lambda_0$ (where $\Lambda_0$ is a constant) in the thermodynamic limit, $L_y\rightarrow \infty$. In contrast, localized states must have $\Lambda^\text{typ}_{L_y}/L_y\rightarrow 0$ in the thermodynamic limit.  In Fig. \ref{fig:timedependenceoflocalizationlength}, we investigate the states in the quasi-energy gap, $\epsilon=\Omega/2$. We pick a disorder strength, $U_0/M=0.9$ for which it appears that the mobility gap has closed. If the disorder strength corresponds to the critical disorder strength of the disorder-induced phase transition, it is expected that these states are extended. With this expectation, we examine the length scale, $\Lambda_N^{\rm typ}$ at disorder strength $U_0/M=0.9$ . Clearly for short times the length scale $\Lambda_N^{\rm typ}$ increases diffusively, $\Lambda_N^{\rm typ}\sim \sqrt{N}$. For long times, this length scale saturates, and the saturation length increases on increasing the width of the strip. The dependence of the saturation length scale on the width of the strip is consistent with either an extended state or a localization length that is much larger than the width of the strip. Within the current numerical accuracy, this is an indication that the localization transition occurs around this disorder strength.
\subsection{Bott index}
 The Bott index is a topological invariant that has been defined by Loring and Hastings\cite{hastingsLoringEPL2010} for disordered systems with broken time-reversal symmetry. For time-independent Hamiltonians, it has also been shown that this index is equivalent to the Hall conductivity of the sample, i.e. the Chern number. We generalize this definition to obtain the disorder-averaged Bott index in periodically driven Hamiltonians. In this section, we elaborate on the prescription to obtain the Bott index for the system under consideration.

 For driven systems, we start with the truncated Floquet Hamiltonian, $H^F$, defined on a lattice with periodic boundary conditions (torus in real space)
 . The coordinates of the lattice points allow us to define two diagonal matrices $X_{ij}=x\delta_{ij}$ and $Y=y\delta_{ij}$, where, $(i,j)$ represents two different sites on the lattice. The two corresponding unitary matrices are, $   U_X= \exp(i2\pi X/L_x)$ and,  $ U_Y= \exp(i2\pi Y/L_y)$. The Bott index for a given Floquet band is obtained from the wavefunctions of the states in this band. Given the quasi-energies are bounded, $\epsilon_l<\epsilon<\epsilon_h$, the Bott index is an integer, that counts the difference in the number of edge states at $\epsilon_l$ and $\epsilon_h$. For our purposes, we set $\epsilon_l=0$. The projector on to this band of Floquet eigenstates, $P$, can be used to project the unitary matrices, to give almost unitary matrices,
    \begin{equation}
      \tilde{U}_{X,Y}= PU_{X,Y}P.
  \end{equation}
 It was shown by Loring and Hastings\cite{hastingsLoringEPL2010}, that these almost unitary matrices, $\tilde{U}^\dagger_{X,Y} \tilde{U}^{}_{X,Y}\approx 1$, are also almost commuting. For brevity, we skip the exact mathematical definitions of almost unitary and almost commuting matrices\cite{hastingsLoringEPL2010}. The Bott index is a measure of commutativity of these projected unitary matrices, $\tilde{U}_{X,Y}$, and is quantized to integers. Explicitly,
\begin{equation}
 C_b=\frac{1}{2\pi}{\rm Im}\left[{\rm Tr}\left\{\ln\left(\tilde{U}^{}_{Y}\tilde{U}^{}_{X}\tilde{U}^{\dag}_{Y} \tilde{U}^{\dag}_{X} \right)\right\}\right].
\end{equation}
This index has been shown to be equivalent to the Kubo formula for the Hall conductivity. The disorder-averaged Bott index at a given quasienergy is then obtained by averaging this index over a large number of different disorder configuration. While the index is an integer for every particular disorder configuration, there is no such requirement on the average Bott index. In fact, through a topological phase transition due to disorder, the average index smoothly changes from one integer to another.
\section{Quantum wells in circularly polarized light}
In this section we provide details for the quantum-well model in the presence of  polarized radiation. We show that in the presence of elliptically polarized light, it is possible to induce insulating band structures with co-propagating edge modes (Chern number $=2$). In the following, we outline the key steps to show that the quasi-energy bandstructure is non-trivial.

In the presence of elliptically polarized light, the time reversal symmetry is explicitly broken.  Through the rest of the section, we will only focus on only one block (upper) of the BHZ model as defined in Eq. (\ref{eq:H_rad}) (with $V=0$). The results can be generalized to the lower block by an appropriate time-reversal operation on the lower block. Let us introduce polarized light through a Peierls substitution of a time-dependent gauge field ${\bf A}\equiv (\phi_x,\phi_y)=\left(A_x \sin(\Omega t),A_y \cos(\Omega t)\right)$,
\begin{eqnarray}
 k_x &\rightarrow& k_x-\phi_x,\\
 k_y &\rightarrow& k_y-\phi_y.
\end{eqnarray}
The Hamiltonian transforms under this substitution, ${\bf d}({\bf k})\cdot{\boldsymbol \sigma}\rightarrow{\bf d}({\bf k}-{\bf A})\cdot {\boldsymbol \sigma}\equiv \tilde{{\bf d}}\cdot{\boldsymbol \sigma}$, and the individual components of the vector, $\tilde{\bf d}\equiv \left(\tilde{d}_x,\tilde{d}_y,\tilde{d}_z\right)$, are,
\begin{eqnarray}
\tilde{d}_x &=& A\left[\sin k_x\cos \phi_x -\cos k_x \sin \phi_x \right]\nonumber\\
\tilde{d}_y &=& A\left[\sin k_y\cos \phi_y -\cos k_y \sin \phi_y \right]\label{eq:circpol_d_def}\\
\tilde{d}_z &=& M-2B[2-(\cos k_x \cos \phi_x +\sin k_x \sin \phi_x  \nonumber\\
& &\ \ \ \ \ \ \ \ \ \ \ \ \ \ \ \ +\cos k_y \cos \phi_y +\sin k_y \sin \phi_y )]\nonumber
\end{eqnarray}
In the numerical simulations, we define $\tilde{\bf d}({\bf k})$ as shown in (\ref{eq:circpol_d_def}). The stroboscopic time-evolution operator, $U(T)$, is obtained  explicitly through a time-ordered integration of the infinitesimal time-evolution operator,which in turn is obtained from exponentiating the time-dependent Hamiltonian given by $\tilde{{\bf d}}(t)\cdot{\boldsymbol \sigma}$.

 Let us consider the perturbative limit for the radiation field to make progress analytically. Setting, $\cos(\phi_{x,y})\approx 1$ and $\sin(\phi_{x,y})\approx \phi_{x,y}$, the gauge-potential as a perturbation to the original Hamiltonian becomes,
\begin{eqnarray}
H= {\bf d}\cdot{\boldsymbol \sigma} +\mathcal{V}\cdot {\boldsymbol \sigma}e^{i\Omega t}+\mathcal{V}^\dagger \cdot {\boldsymbol \sigma}e^{-i\Omega t},
\end{eqnarray}
where we have defined,
\begin{eqnarray}
\mathcal{V}=\left(\frac{iAA_x }{2}\cos k_x,-\frac{AA_y}{2}\cos k_y,B(A_y\sin k_y-iA_x \sin k_x)\right)\rule[-1.5em]{0pt}{0pt}\nonumber\\
\end{eqnarray}
\begin{figure}
\includegraphics[width=0.7\linewidth]{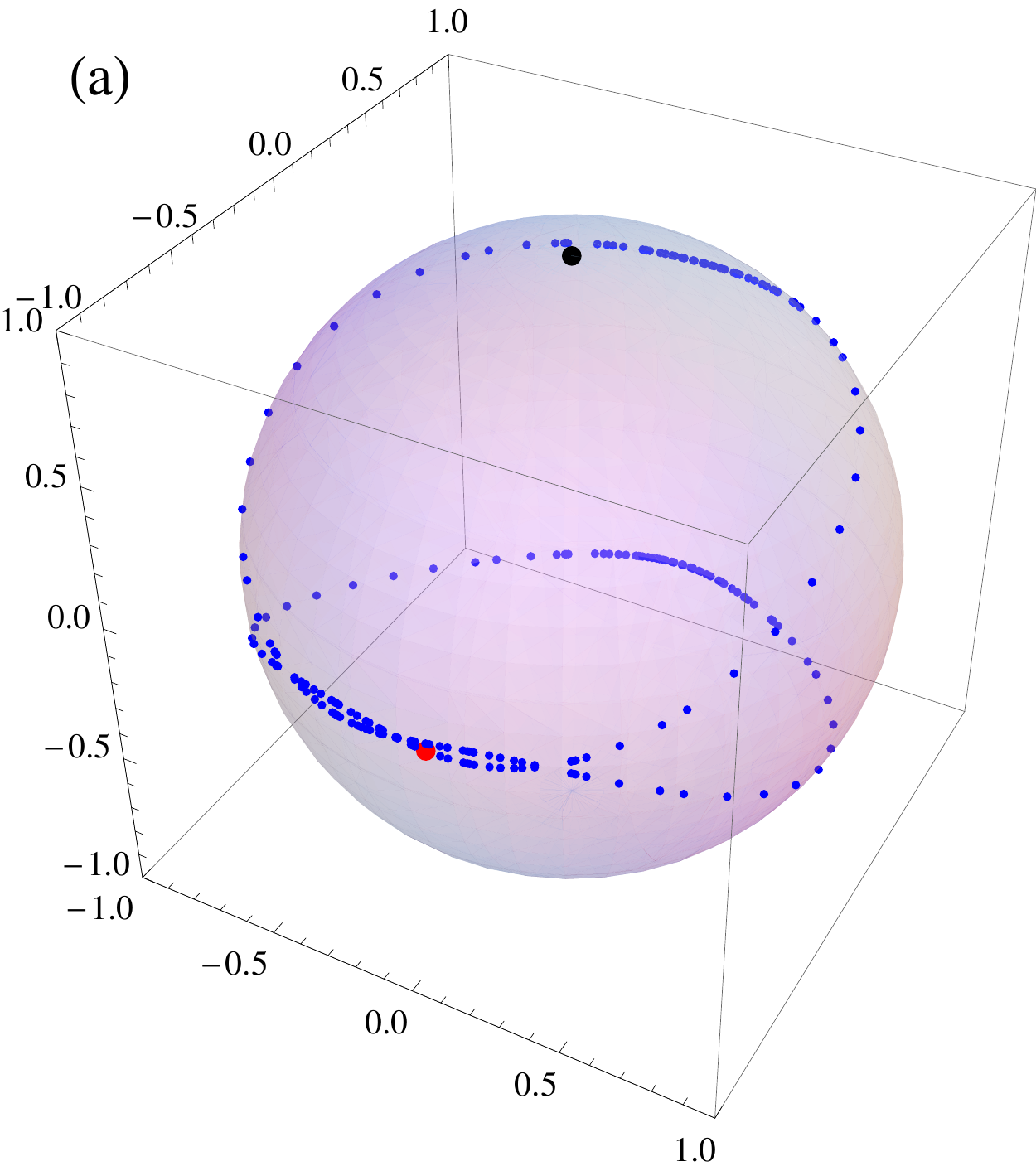}
\caption{Winding of ${\bf V}_\perp({\bf k})$ on the unit sphere [See Eqs. (\ref{eq:Vperp-parametric}-\ref{eq:V-imaginary})] as the momentum vector, ${\bf k}$ varies along the resonance circle [defined in Eq.(\ref{eq:resonancecondition})]. The black point indicates the north pole, $(0,0,1)$ and the red point indicates $\hat{V}_\perp({\bf k}_0)$ with ${\bf k}_0=(1,0)$. }
\label{fig:resonancewinding}
\end{figure}
As discussed in section II, the quasi-energy gap at resonance is governed by the matrix element, $\left|\langle\psi_+|\mathcal{V}\cdot{\boldsymbol \sigma}|\psi_-\rangle\right|$, where $|\psi_\pm\rangle$ are the eigenstates of the unperturbed Hamiltonian. In the original basis, the eigenstates are,
\begin{eqnarray}
|\psi_+\rangle &=& \cos\frac{\theta}{2}|\uparrow\rangle+\sin \frac{\theta}{2}e^{i\phi}|\downarrow\rangle,\\
|\psi_-\rangle &=& \sin\frac{\theta}{2}|\uparrow\rangle-\cos \frac{\theta}{2}e^{i\phi}|\downarrow\rangle,
\end{eqnarray}
where ${\bf d}({\bf k})\equiv |{\bf d}|(\sin\theta\cos\phi,\sin\theta\sin\phi,\cos\theta)$, and the original basis is defined as, $\{|\uparrow\rangle,|\downarrow\rangle\}$. The winding around the north pole, ${\bf V}_\perp$ (see Eq. \ref{eq:Vperp}) is necessarily related to the Chern number of the Floquet bands. In the original basis, $V_\perp$ is defined as,
\begin{equation}
{\bf V}_\perp\cdot {\boldsymbol \sigma}\equiv \langle\psi_+|\mathcal{V}\cdot{\boldsymbol \sigma}|\psi_-\rangle|\psi_+\rangle\langle\psi_-|+{\rm h.c.}\label{eq:Vdef-1}
\end{equation}
In order to obtain ${\bf V}_\perp$ from $\mathcal{V}$, we make use of the following identities,
\begin{eqnarray}
|\psi_+\rangle\langle\psi_-|&=& \begin{pmatrix}
\cos\frac{\theta}{2}\sin\frac{\theta}{2} && -\cos ^2\frac{\theta}{2}e^{-i\phi}\\
\sin^2\frac{\theta}{2}e^{i\phi}&&-\cos\frac{\theta}{2}\sin\frac{\theta}{2}
\end{pmatrix},\\
\langle\psi_+|\sigma_x|\psi_-\rangle &=& -\cos\theta \cos\phi-i\sin\phi, \\
\langle\psi_+|\sigma_y|\psi_-\rangle &=& -\cos\theta \sin\phi+i\cos\phi,\\
\langle\psi_+|\sigma_z|\psi_-\rangle &=& \sin\theta.
\end{eqnarray}

Using the above identities in Eq. (\ref{eq:Vdef-1}), and rewriting the vector, $\langle\psi_+|\mathcal{V}|\psi_-\rangle=\mathcal{V}_R+i\mathcal{V}_I$,
\begin{eqnarray}
{\bf V}_\perp&=&\mathcal{V}_R(-\cos\theta\cos\phi,-\cos\theta\sin\phi,\sin\theta)\nonumber\\
&&+\mathcal{V}_I(-\sin\phi,\cos\phi,0),\label{eq:Vperp-parametric}\\
{\rm with,}\nonumber\\
\mathcal{V}_R&=& \frac{AA_x}{2}\sin\phi\cos k_x+\frac{AA_y}{2}\cos\theta \sin \phi\cos k_y\nonumber\\
&&+B A_y \sin k_y \sin\theta,\label{eq:V-real}\\
\mathcal{V}_I&=& -\frac{AA_x}{2}\cos\theta\cos\phi\cos k_x-\frac{AA_y}{2}\cos\phi\cos k_y\nonumber\\
&&-B A_x \sin k_x \sin\theta. \label{eq:V-imaginary}
\end{eqnarray}
\begin{figure} h
\includegraphics[width=\linewidth]{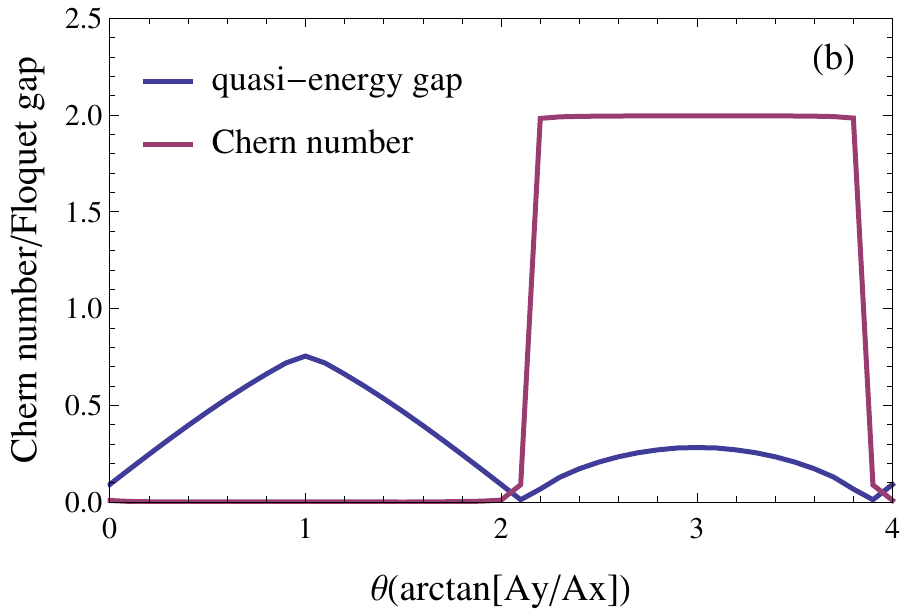}
\caption{We show the dependence on polarization of light parameterized by, $\theta=\arctan(A_y/A_x)$ of the Chern number and the gap in the quasi-energy bandstructure of the upper block.}
\label{fig:chernnumberandgaps}
\end{figure}
Note that the angles $\theta$ and $\phi$ are obtained from the definitions of ${\bf d}({\bf k})$. We are interested in ${\bf V}_\perp$ along the resonance circle given by $|{\bf d}({\bf k})|=\frac{\Omega}{2}$. Let us now make a second approximation, and take the limit, $k_{x,y}\rightarrow 0$, and expand to second order in $k_{x,y}$. In this limit, we denote the resonance circle by a constant parameter $k_r=\sqrt{k_x^2+k_y^2}$. The definitions in Eqs. (\ref{eq:Vperp-parametric}), (\ref{eq:V-real}) and (\ref{eq:V-imaginary}) reduce to,
\begin{eqnarray}
{\bf V}_\perp &=&\mathcal{V}_R\left(-\frac{M-Bk_r^2}{|{\bf d}|k_r}k_x,-\frac{M-Bk_r^2}{|{\bf d}|k_r}k_y,\frac{Ak_r}{|{\bf d}|}\right)\nonumber\\
&&+\mathcal{V}_I\left(-\frac{A}{k_r}k_y,\frac{A}{k_r}k_x,0\right)\\
\mathcal{V}_R&=&\left[\frac{AA_x}{2k_r}+\frac{AA_y(M-Bk_r^2)}{2k_r|{\bf d}|}+\frac{BAA_yk_r}{|{\bf d}|}\right] k_y,\ \ \ \ \\
\mathcal{V}_I&=&\left[-\frac{AA_x(M-Bk_r^2)}{2k_r|{\bf d}|}-\frac{AA_y}{2k_r}-\frac{BAA_x k_r}{|{\bf d}|}\right] k_x,\ \ \ \
\end{eqnarray}
where, $|{\bf d}|=\Omega/2$ is a constant.

The Chern number of the bands obtained from irradiation of the quantum wells with circularly polarized light is $\pm 2$. Figure (\ref{fig:resonancewinding}) shows the winding of the vector ${\bf V}_\perp$ along the resonance circle. Clearly, it winds twice around the north pole, which is consistent with a Chern number $=\pm2$. The Chern-number of the band also depends on the polarization of the incident light as shown in Fig. (\ref{fig:chernnumberandgaps}). The approximate value of the Chern number for a given polarization angle, $\theta$, is obtained by computing $C$ defined in Eq. (\ref{eq:Chernnumberformula}). The quasi-energy gap in the band structure depends on the incident polarization of the radiation, given by $|{\bf V}_\perp|$. Clearly, the gap closes as function of $\theta$, when a transition happens from a topological ($\mathcal{C}=2$) to the  trivial ($\mathcal{C}=0$) phase.

 In order to realize a FTAI in these systems, we choose a polarization angle that is close to the boundary and on the trivial side of the transition. Disorder renormalizes the position of the boundary of this transition.
\bibliographystyle{apsrev}
\bibliography{disorderedfloqrefs}

\begin{thebibliography}{65}
\expandafter\ifx\csname natexlab\endcsname\relax\def\natexlab#1{#1}\fi
\expandafter\ifx\csname bibnamefont\endcsname\relax
  \def\bibnamefont#1{#1}\fi
\expandafter\ifx\csname bibfnamefont\endcsname\relax
  \def\bibfnamefont#1{#1}\fi
\expandafter\ifx\csname citenamefont\endcsname\relax
  \def\citenamefont#1{#1}\fi
\expandafter\ifx\csname url\endcsname\relax
  \def\url#1{\texttt{#1}}\fi
\expandafter\ifx\csname urlprefix\endcsname\relax\def\urlprefix{URL }\fi
\providecommand{\bibinfo}[2]{#2}
\providecommand{\eprint}[2][]{\url{#2}}

\bibitem[{\citenamefont{Lindner et~al.}(2011)\citenamefont{Lindner, Refael, and
  Galitski}}]{Lindner2011}
\bibinfo{author}{\bibfnamefont{N.~H.} \bibnamefont{Lindner}},
  \bibinfo{author}{\bibfnamefont{G.}~\bibnamefont{Refael}}, \bibnamefont{and}
  \bibinfo{author}{\bibfnamefont{V.}~\bibnamefont{Galitski}},
  \bibinfo{journal}{Nat. Phys.} \textbf{\bibinfo{volume}{7}},
  \bibinfo{pages}{490} (\bibinfo{year}{2011}).

\bibitem[{\citenamefont{Oka and Aoki}(2009)}]{Oka2009}
\bibinfo{author}{\bibfnamefont{T.}~\bibnamefont{Oka}} \bibnamefont{and}
  \bibinfo{author}{\bibfnamefont{H.}~\bibnamefont{Aoki}},
  \bibinfo{journal}{Phys. Rev. B} \textbf{\bibinfo{volume}{79}},
  \bibinfo{pages}{081406} (\bibinfo{year}{2009}).

\bibitem[{\citenamefont{Kitagawa et~al.}(2011)\citenamefont{Kitagawa, Oka,
  Brataas, Fu, and Demler}}]{Kitagawa2011}
\bibinfo{author}{\bibfnamefont{T.}~\bibnamefont{Kitagawa}},
  \bibinfo{author}{\bibfnamefont{T.}~\bibnamefont{Oka}},
  \bibinfo{author}{\bibfnamefont{A.}~\bibnamefont{Brataas}},
  \bibinfo{author}{\bibfnamefont{L.}~\bibnamefont{Fu}}, \bibnamefont{and}
  \bibinfo{author}{\bibfnamefont{E.}~\bibnamefont{Demler}},
  \bibinfo{journal}{Phys. Rev. B} \textbf{\bibinfo{volume}{84}},
  \bibinfo{pages}{235108} (\bibinfo{year}{2011}).

\bibitem[{\citenamefont{Jiang et~al.}(2011)\citenamefont{Jiang, Kitagawa,
  Alicea, Akhmerov, Pekker, Refael, Cirac, Demler, Lukin, and
  Zoller}}]{Jiang2011}
\bibinfo{author}{\bibfnamefont{L.}~\bibnamefont{Jiang}},
  \bibinfo{author}{\bibfnamefont{T.}~\bibnamefont{Kitagawa}},
  \bibinfo{author}{\bibfnamefont{J.}~\bibnamefont{Alicea}},
  \bibinfo{author}{\bibfnamefont{A.~R.} \bibnamefont{Akhmerov}},
  \bibinfo{author}{\bibfnamefont{D.}~\bibnamefont{Pekker}},
  \bibinfo{author}{\bibfnamefont{G.}~\bibnamefont{Refael}},
  \bibinfo{author}{\bibfnamefont{J.~I.} \bibnamefont{Cirac}},
  \bibinfo{author}{\bibfnamefont{E.}~\bibnamefont{Demler}},
  \bibinfo{author}{\bibfnamefont{M.~D.} \bibnamefont{Lukin}}, \bibnamefont{and}
  \bibinfo{author}{\bibfnamefont{P.}~\bibnamefont{Zoller}},
  \bibinfo{journal}{Phys. Rev. Lett.} \textbf{\bibinfo{volume}{106}},
  \bibinfo{pages}{220402} (\bibinfo{year}{2011}).

\bibitem[{\citenamefont{Lindner et~al.}(2013)\citenamefont{Lindner, Bergman,
  Refael, and Galitski}}]{Lindner2013}
\bibinfo{author}{\bibfnamefont{N.~H.} \bibnamefont{Lindner}},
  \bibinfo{author}{\bibfnamefont{D.~L.} \bibnamefont{Bergman}},
  \bibinfo{author}{\bibfnamefont{G.}~\bibnamefont{Refael}}, \bibnamefont{and}
  \bibinfo{author}{\bibfnamefont{V.}~\bibnamefont{Galitski}},
  \bibinfo{journal}{Phys. Rev. B} \textbf{\bibinfo{volume}{87}},
  \bibinfo{pages}{235131} (\bibinfo{year}{2013}).

\bibitem[{\citenamefont{Katan and Podolsky}(2013)}]{Podolsky2013}
\bibinfo{author}{\bibfnamefont{Y.~T.} \bibnamefont{Katan}} \bibnamefont{and}
  \bibinfo{author}{\bibfnamefont{D.}~\bibnamefont{Podolsky}},
  \bibinfo{journal}{Phys. Rev. Lett.} \textbf{\bibinfo{volume}{110}},
  \bibinfo{pages}{016802} (\bibinfo{year}{2013}).

\bibitem[{\citenamefont{Usaj et~al.}(2014)\citenamefont{Usaj, Perez-Piskunow,
  Foa~Torres, and Balseiro}}]{TorresPRB2014}
\bibinfo{author}{\bibfnamefont{G.}~\bibnamefont{Usaj}},
  \bibinfo{author}{\bibfnamefont{P.~M.} \bibnamefont{Perez-Piskunow}},
  \bibinfo{author}{\bibfnamefont{L.~E.~F.} \bibnamefont{Foa~Torres}},
  \bibnamefont{and} \bibinfo{author}{\bibfnamefont{C.~A.}
  \bibnamefont{Balseiro}}, \bibinfo{journal}{Phys. Rev. B}
  \textbf{\bibinfo{volume}{90}}, \bibinfo{pages}{115423}
  (\bibinfo{year}{2014}).

\bibitem[{\citenamefont{Gu et~al.}(2011)\citenamefont{Gu, Fertig, Arovas, and
  Auerbach}}]{Gu11}
\bibinfo{author}{\bibfnamefont{Z.}~\bibnamefont{Gu}},
  \bibinfo{author}{\bibfnamefont{H.~A.} \bibnamefont{Fertig}},
  \bibinfo{author}{\bibfnamefont{D.~P.} \bibnamefont{Arovas}},
  \bibnamefont{and} \bibinfo{author}{\bibfnamefont{A.}~\bibnamefont{Auerbach}},
  \bibinfo{journal}{Phys. Rev. Lett.} \textbf{\bibinfo{volume}{107}},
  \bibinfo{pages}{216601} (\bibinfo{year}{2011}).

\bibitem[{\citenamefont{Liu et~al.}(2013)\citenamefont{Liu, Levchenko, and
  Baranger}}]{LiuPRL2013}
\bibinfo{author}{\bibfnamefont{D.~E.} \bibnamefont{Liu}},
  \bibinfo{author}{\bibfnamefont{A.}~\bibnamefont{Levchenko}},
  \bibnamefont{and} \bibinfo{author}{\bibfnamefont{H.~U.}
  \bibnamefont{Baranger}}, \bibinfo{journal}{Phys. Rev. Lett.}
  \textbf{\bibinfo{volume}{111}}, \bibinfo{pages}{047002}
  (\bibinfo{year}{2013}).

\bibitem[{\citenamefont{Li et~al.}(2014)\citenamefont{Li, Kundu, Zhong, and
  Seradjeh}}]{KunduPRB2014}
\bibinfo{author}{\bibfnamefont{Y.}~\bibnamefont{Li}},
  \bibinfo{author}{\bibfnamefont{A.}~\bibnamefont{Kundu}},
  \bibinfo{author}{\bibfnamefont{F.}~\bibnamefont{Zhong}}, \bibnamefont{and}
  \bibinfo{author}{\bibfnamefont{B.}~\bibnamefont{Seradjeh}},
  \bibinfo{journal}{Phys. Rev. B} \textbf{\bibinfo{volume}{90}},
  \bibinfo{pages}{121401} (\bibinfo{year}{2014}).

\bibitem[{\citenamefont{Rudner et~al.}(2013)\citenamefont{Rudner, Lindner,
  Berg, and Levin}}]{Rudner2013}
\bibinfo{author}{\bibfnamefont{M.~S.} \bibnamefont{Rudner}},
  \bibinfo{author}{\bibfnamefont{N.~H.} \bibnamefont{Lindner}},
  \bibinfo{author}{\bibfnamefont{E.}~\bibnamefont{Berg}}, \bibnamefont{and}
  \bibinfo{author}{\bibfnamefont{M.}~\bibnamefont{Levin}},
  \bibinfo{journal}{Phys. Rev. X} \textbf{\bibinfo{volume}{3}},
  \bibinfo{pages}{031005} (\bibinfo{year}{2013}).

\bibitem[{\citenamefont{Titum et~al.}(2016)\citenamefont{Titum, Berg, Rudner,
  Refael, and Lindner}}]{Titumarxiv2015}
\bibinfo{author}{\bibfnamefont{P.}~\bibnamefont{Titum}},
  \bibinfo{author}{\bibfnamefont{E.}~\bibnamefont{Berg}},
  \bibinfo{author}{\bibfnamefont{M.~S.} \bibnamefont{Rudner}},
  \bibinfo{author}{\bibfnamefont{G.}~\bibnamefont{Refael}}, \bibnamefont{and}
  \bibinfo{author}{\bibfnamefont{N.~H.} \bibnamefont{Lindner}},
  \bibinfo{journal}{Phys. Rev. X} \textbf{\bibinfo{volume}{6}},
  \bibinfo{pages}{021013} (\bibinfo{year}{2016}).

\bibitem[{\citenamefont{Else et~al.}(2016)\citenamefont{Else, Bauer, and
  Nayak}}]{Else_bauer_Nayak_PRL2016}
\bibinfo{author}{\bibfnamefont{D.~V.} \bibnamefont{Else}},
  \bibinfo{author}{\bibfnamefont{B.}~\bibnamefont{Bauer}}, \bibnamefont{and}
  \bibinfo{author}{\bibfnamefont{C.}~\bibnamefont{Nayak}},
  \bibinfo{journal}{Phys. Rev. Lett.} \textbf{\bibinfo{volume}{117}},
  \bibinfo{pages}{090402} (\bibinfo{year}{2016}).

\bibitem[{\citenamefont{Else et~al.}(2017)\citenamefont{Else, Bauer, and
  Nayak}}]{Else_Bauer2016}
\bibinfo{author}{\bibfnamefont{D.~V.} \bibnamefont{Else}},
  \bibinfo{author}{\bibfnamefont{B.}~\bibnamefont{Bauer}}, \bibnamefont{and}
  \bibinfo{author}{\bibfnamefont{C.}~\bibnamefont{Nayak}},
  \bibinfo{journal}{Phys. Rev. X} \textbf{\bibinfo{volume}{7}},
  \bibinfo{pages}{011026} (\bibinfo{year}{2017}).

\bibitem[{\citenamefont{Po et~al.}(2016)\citenamefont{Po, Fidkowski, Morimoto,
  Potter, and Vishwanath}}]{Po_Vishwanath2016}
\bibinfo{author}{\bibfnamefont{H.~C.} \bibnamefont{Po}},
  \bibinfo{author}{\bibfnamefont{L.}~\bibnamefont{Fidkowski}},
  \bibinfo{author}{\bibfnamefont{T.}~\bibnamefont{Morimoto}},
  \bibinfo{author}{\bibfnamefont{A.~C.} \bibnamefont{Potter}},
  \bibnamefont{and}
  \bibinfo{author}{\bibfnamefont{A.}~\bibnamefont{Vishwanath}},
  \bibinfo{journal}{Phys. Rev. X} \textbf{\bibinfo{volume}{6}},
  \bibinfo{pages}{041070} (\bibinfo{year}{2016}).

\bibitem[{\citenamefont{Po et~al.}(2017)\citenamefont{Po, Fidkowski, Potter,
  and Vishwanath}}]{Po_Vishwanath2017}
\bibinfo{author}{\bibfnamefont{H.~C.} \bibnamefont{Po}},
  \bibinfo{author}{\bibfnamefont{L.}~\bibnamefont{Fidkowski}},
  \bibinfo{author}{\bibfnamefont{A.~C.} \bibnamefont{Potter}},
  \bibnamefont{and}
  \bibinfo{author}{\bibfnamefont{A.}~\bibnamefont{Vishwanath}},
  \bibinfo{journal}{arXiv:1701.01440}  (\bibinfo{year}{2017}).

\bibitem[{\citenamefont{Harper and Roy}(2017)}]{Harper_Roy2016}
\bibinfo{author}{\bibfnamefont{F.}~\bibnamefont{Harper}} \bibnamefont{and}
  \bibinfo{author}{\bibfnamefont{R.}~\bibnamefont{Roy}},
  \bibinfo{journal}{Phys. Rev. Lett.} \textbf{\bibinfo{volume}{118}},
  \bibinfo{pages}{115301} (\bibinfo{year}{2017}).

\bibitem[{\citenamefont{Khemani et~al.}(2016)\citenamefont{Khemani, Lazarides,
  Moessner, and Sondhi}}]{KhemaniPRL2016}
\bibinfo{author}{\bibfnamefont{V.}~\bibnamefont{Khemani}},
  \bibinfo{author}{\bibfnamefont{A.}~\bibnamefont{Lazarides}},
  \bibinfo{author}{\bibfnamefont{R.}~\bibnamefont{Moessner}}, \bibnamefont{and}
  \bibinfo{author}{\bibfnamefont{S.~L.} \bibnamefont{Sondhi}},
  \bibinfo{journal}{Phys. Rev. Lett.} \textbf{\bibinfo{volume}{116}},
  \bibinfo{pages}{250401} (\bibinfo{year}{2016}).

\bibitem[{\citenamefont{Nathan and Rudner}(2015)}]{Nathan_NJP2015}
\bibinfo{author}{\bibfnamefont{F.}~\bibnamefont{Nathan}} \bibnamefont{and}
  \bibinfo{author}{\bibfnamefont{M.~S.} \bibnamefont{Rudner}},
  \bibinfo{journal}{New Journal of Physics} \textbf{\bibinfo{volume}{17}},
  \bibinfo{pages}{125014} (\bibinfo{year}{2015}).

\bibitem[{\citenamefont{Nathan et~al.}(2016)\citenamefont{Nathan, Rudner,
  Lindner, Berg, and Refael}}]{Nathan_arxiv2016}
\bibinfo{author}{\bibfnamefont{F.}~\bibnamefont{Nathan}},
  \bibinfo{author}{\bibfnamefont{M.~S.} \bibnamefont{Rudner}},
  \bibinfo{author}{\bibfnamefont{N.~H.} \bibnamefont{Lindner}},
  \bibinfo{author}{\bibfnamefont{E.}~\bibnamefont{Berg}}, \bibnamefont{and}
  \bibinfo{author}{\bibfnamefont{G.}~\bibnamefont{Refael}},
  \bibinfo{journal}{arXiv:1610.03590}  (\bibinfo{year}{2016}).

\bibitem[{\citenamefont{Roy and
  Harper}(2016{\natexlab{a}})}]{Roy_Harper_Prb2016}
\bibinfo{author}{\bibfnamefont{R.}~\bibnamefont{Roy}} \bibnamefont{and}
  \bibinfo{author}{\bibfnamefont{F.}~\bibnamefont{Harper}},
  \bibinfo{journal}{Phys. Rev. B} \textbf{\bibinfo{volume}{94}},
  \bibinfo{pages}{125105} (\bibinfo{year}{2016}{\natexlab{a}}).

\bibitem[{\citenamefont{Else and Nayak}(2016)}]{Else_Nayak_PRB2016}
\bibinfo{author}{\bibfnamefont{D.~V.} \bibnamefont{Else}} \bibnamefont{and}
  \bibinfo{author}{\bibfnamefont{C.}~\bibnamefont{Nayak}},
  \bibinfo{journal}{Phys. Rev. B} \textbf{\bibinfo{volume}{93}},
  \bibinfo{pages}{201103} (\bibinfo{year}{2016}).

\bibitem[{\citenamefont{von Keyserlingk and
  Sondhi}(2016{\natexlab{a}})}]{Keyserlingk_PRB2016}
\bibinfo{author}{\bibfnamefont{C.~W.} \bibnamefont{von Keyserlingk}}
  \bibnamefont{and} \bibinfo{author}{\bibfnamefont{S.~L.}
  \bibnamefont{Sondhi}}, \bibinfo{journal}{Phys. Rev. B}
  \textbf{\bibinfo{volume}{93}}, \bibinfo{pages}{245145}
  (\bibinfo{year}{2016}{\natexlab{a}}).

\bibitem[{\citenamefont{von Keyserlingk and
  Sondhi}(2016{\natexlab{b}})}]{Keyserlingk_PRB2016-II}
\bibinfo{author}{\bibfnamefont{C.~W.} \bibnamefont{von Keyserlingk}}
  \bibnamefont{and} \bibinfo{author}{\bibfnamefont{S.~L.}
  \bibnamefont{Sondhi}}, \bibinfo{journal}{Phys. Rev. B}
  \textbf{\bibinfo{volume}{93}}, \bibinfo{pages}{245146}
  (\bibinfo{year}{2016}{\natexlab{b}}).

\bibitem[{\citenamefont{von Keyserlingk et~al.}(2016)\citenamefont{von
  Keyserlingk, Khemani, and Sondhi}}]{Keyserlingk_PRB2016-III}
\bibinfo{author}{\bibfnamefont{C.~W.} \bibnamefont{von Keyserlingk}},
  \bibinfo{author}{\bibfnamefont{V.}~\bibnamefont{Khemani}}, \bibnamefont{and}
  \bibinfo{author}{\bibfnamefont{S.~L.} \bibnamefont{Sondhi}},
  \bibinfo{journal}{Phys. Rev. B} \textbf{\bibinfo{volume}{94}},
  \bibinfo{pages}{085112} (\bibinfo{year}{2016}).

\bibitem[{\citenamefont{Potter et~al.}(2016)\citenamefont{Potter, Morimoto, and
  Vishwanath}}]{Potter_PRX2016}
\bibinfo{author}{\bibfnamefont{A.~C.} \bibnamefont{Potter}},
  \bibinfo{author}{\bibfnamefont{T.}~\bibnamefont{Morimoto}}, \bibnamefont{and}
  \bibinfo{author}{\bibfnamefont{A.}~\bibnamefont{Vishwanath}},
  \bibinfo{journal}{Phys. Rev. X} \textbf{\bibinfo{volume}{6}},
  \bibinfo{pages}{041001} (\bibinfo{year}{2016}).

\bibitem[{\citenamefont{Roy and Harper}(2016{\natexlab{b}})}]{Roy_Harper2016}
\bibinfo{author}{\bibfnamefont{R.}~\bibnamefont{Roy}} \bibnamefont{and}
  \bibinfo{author}{\bibfnamefont{F.}~\bibnamefont{Harper}},
  \bibinfo{journal}{arXiv:1603.06944}  (\bibinfo{year}{2016}{\natexlab{b}}).

\bibitem[{\citenamefont{Potter and Morimoto}(2017)}]{Potter_Morimoto_2016}
\bibinfo{author}{\bibfnamefont{A.~C.} \bibnamefont{Potter}} \bibnamefont{and}
  \bibinfo{author}{\bibfnamefont{T.}~\bibnamefont{Morimoto}},
  \bibinfo{journal}{Phys. Rev. B} \textbf{\bibinfo{volume}{95}},
  \bibinfo{pages}{155126} (\bibinfo{year}{2017}).

\bibitem[{\citenamefont{Wang et~al.}(2013)\citenamefont{Wang, Steinberg,
  Jarillo-Herrero, and Gedik}}]{Wang2013}
\bibinfo{author}{\bibfnamefont{Y.~H.} \bibnamefont{Wang}},
  \bibinfo{author}{\bibfnamefont{H.}~\bibnamefont{Steinberg}},
  \bibinfo{author}{\bibfnamefont{P.}~\bibnamefont{Jarillo-Herrero}},
  \bibnamefont{and} \bibinfo{author}{\bibfnamefont{N.}~\bibnamefont{Gedik}},
  \bibinfo{journal}{Science} \textbf{\bibinfo{volume}{342}},
  \bibinfo{pages}{453} (\bibinfo{year}{2013}).

\bibitem[{\citenamefont{Mahmood et~al.}(2016)\citenamefont{Mahmood, Chan,
  Alpichshev, Gardner, Lee, Lee, and Gedik}}]{Gedik_Floquet-2}
\bibinfo{author}{\bibfnamefont{F.}~\bibnamefont{Mahmood}},
  \bibinfo{author}{\bibfnamefont{C.-K.} \bibnamefont{Chan}},
  \bibinfo{author}{\bibfnamefont{Z.}~\bibnamefont{Alpichshev}},
  \bibinfo{author}{\bibfnamefont{D.}~\bibnamefont{Gardner}},
  \bibinfo{author}{\bibfnamefont{Y.}~\bibnamefont{Lee}},
  \bibinfo{author}{\bibfnamefont{P.~A.} \bibnamefont{Lee}}, \bibnamefont{and}
  \bibinfo{author}{\bibfnamefont{N.}~\bibnamefont{Gedik}},
  \bibinfo{journal}{Nature Physics} \textbf{\bibinfo{volume}{12}},
  \bibinfo{pages}{306} (\bibinfo{year}{2016}).

\bibitem[{\citenamefont{Rechtsman et~al.}(2013)\citenamefont{Rechtsman, Zeuner,
  Plotnik, Lumer, Podolsky, Dreisow, Nolte, Segev, and
  Szameit}}]{Rechtsman2013}
\bibinfo{author}{\bibfnamefont{M.~C.} \bibnamefont{Rechtsman}},
  \bibinfo{author}{\bibfnamefont{J.~M.} \bibnamefont{Zeuner}},
  \bibinfo{author}{\bibfnamefont{Y.}~\bibnamefont{Plotnik}},
  \bibinfo{author}{\bibfnamefont{Y.}~\bibnamefont{Lumer}},
  \bibinfo{author}{\bibfnamefont{D.}~\bibnamefont{Podolsky}},
  \bibinfo{author}{\bibfnamefont{F.}~\bibnamefont{Dreisow}},
  \bibinfo{author}{\bibfnamefont{S.}~\bibnamefont{Nolte}},
  \bibinfo{author}{\bibfnamefont{M.}~\bibnamefont{Segev}}, \bibnamefont{and}
  \bibinfo{author}{\bibfnamefont{A.}~\bibnamefont{Szameit}},
  \bibinfo{journal}{Nature} \textbf{\bibinfo{volume}{496}},
  \bibinfo{pages}{196} (\bibinfo{year}{2013}).

\bibitem[{\citenamefont{Jotzu et~al.}(2014)\citenamefont{Jotzu, Messer,
  Desbuquois, Lebrat, Uehlinger, Greif, and Esslinger}}]{Jotzu2014}
\bibinfo{author}{\bibfnamefont{G.}~\bibnamefont{Jotzu}},
  \bibinfo{author}{\bibfnamefont{M.}~\bibnamefont{Messer}},
  \bibinfo{author}{\bibfnamefont{R.}~\bibnamefont{Desbuquois}},
  \bibinfo{author}{\bibfnamefont{M.}~\bibnamefont{Lebrat}},
  \bibinfo{author}{\bibfnamefont{T.}~\bibnamefont{Uehlinger}},
  \bibinfo{author}{\bibfnamefont{D.}~\bibnamefont{Greif}}, \bibnamefont{and}
  \bibinfo{author}{\bibfnamefont{T.}~\bibnamefont{Esslinger}},
  \bibinfo{journal}{Nature} \textbf{\bibinfo{volume}{515}},
  \bibinfo{pages}{237} (\bibinfo{year}{2014}).

\bibitem[{\citenamefont{Foa~Torres et~al.}(2014)\citenamefont{Foa~Torres,
  Perez-Piskunow, Balseiro, and Usaj}}]{TorresPRL2014}
\bibinfo{author}{\bibfnamefont{L.~E.~F.} \bibnamefont{Foa~Torres}},
  \bibinfo{author}{\bibfnamefont{P.~M.} \bibnamefont{Perez-Piskunow}},
  \bibinfo{author}{\bibfnamefont{C.~A.} \bibnamefont{Balseiro}},
  \bibnamefont{and} \bibinfo{author}{\bibfnamefont{G.}~\bibnamefont{Usaj}},
  \bibinfo{journal}{Phys. Rev. Lett.} \textbf{\bibinfo{volume}{113}},
  \bibinfo{pages}{266801} (\bibinfo{year}{2014}).

\bibitem[{\citenamefont{Kundu and Seradjeh}(2013)}]{Kundu2013}
\bibinfo{author}{\bibfnamefont{A.}~\bibnamefont{Kundu}} \bibnamefont{and}
  \bibinfo{author}{\bibfnamefont{B.}~\bibnamefont{Seradjeh}},
  \bibinfo{journal}{Phys. Rev. Lett.} \textbf{\bibinfo{volume}{111}},
  \bibinfo{pages}{136402} (\bibinfo{year}{2013}).

\bibitem[{\citenamefont{Farrell and Pereg-Barnea}(2016)}]{Farrell2015a}
\bibinfo{author}{\bibfnamefont{A.}~\bibnamefont{Farrell}} \bibnamefont{and}
  \bibinfo{author}{\bibfnamefont{T.}~\bibnamefont{Pereg-Barnea}},
  \bibinfo{journal}{Phys. Rev. B} \textbf{\bibinfo{volume}{93}},
  \bibinfo{pages}{045121} (\bibinfo{year}{2016}).

\bibitem[{\citenamefont{Farrell and Pereg-Barnea}(2015)}]{Farrell2015b}
\bibinfo{author}{\bibfnamefont{A.}~\bibnamefont{Farrell}} \bibnamefont{and}
  \bibinfo{author}{\bibfnamefont{T.}~\bibnamefont{Pereg-Barnea}},
  \bibinfo{journal}{Phys. Rev. Lett.} \textbf{\bibinfo{volume}{115}},
  \bibinfo{pages}{106403} (\bibinfo{year}{2015}).

\bibitem[{\citenamefont{Wang et~al.}(2014)\citenamefont{Wang, Sun, and
  Xie}}]{XiePRB2014}
\bibinfo{author}{\bibfnamefont{P.}~\bibnamefont{Wang}},
  \bibinfo{author}{\bibfnamefont{Q.-f.} \bibnamefont{Sun}}, \bibnamefont{and}
  \bibinfo{author}{\bibfnamefont{X.~C.} \bibnamefont{Xie}},
  \bibinfo{journal}{Phys. Rev. B} \textbf{\bibinfo{volume}{90}},
  \bibinfo{pages}{155407} (\bibinfo{year}{2014}).

\bibitem[{\citenamefont{Seetharam et~al.}(2015)\citenamefont{Seetharam, Bardyn,
  Lindner, Rudner, and Refael}}]{Seetharam2015}
\bibinfo{author}{\bibfnamefont{K.~I.} \bibnamefont{Seetharam}},
  \bibinfo{author}{\bibfnamefont{C.-E.} \bibnamefont{Bardyn}},
  \bibinfo{author}{\bibfnamefont{N.~H.} \bibnamefont{Lindner}},
  \bibinfo{author}{\bibfnamefont{M.~S.} \bibnamefont{Rudner}},
  \bibnamefont{and} \bibinfo{author}{\bibfnamefont{G.}~\bibnamefont{Refael}},
  \bibinfo{journal}{Phys. Rev. X} \textbf{\bibinfo{volume}{5}},
  \bibinfo{pages}{041050} (\bibinfo{year}{2015}).

\bibitem[{\citenamefont{Dehghani et~al.}(2014)\citenamefont{Dehghani, Oka, and
  Mitra}}]{Dehghani2014}
\bibinfo{author}{\bibfnamefont{H.}~\bibnamefont{Dehghani}},
  \bibinfo{author}{\bibfnamefont{T.}~\bibnamefont{Oka}}, \bibnamefont{and}
  \bibinfo{author}{\bibfnamefont{A.}~\bibnamefont{Mitra}},
  \bibinfo{journal}{Phys. Rev. B} \textbf{\bibinfo{volume}{90}},
  \bibinfo{pages}{195429} (\bibinfo{year}{2014}).

\bibitem[{\citenamefont{Dehghani et~al.}(2015)\citenamefont{Dehghani, Oka, and
  Mitra}}]{Dehghani2014b}
\bibinfo{author}{\bibfnamefont{H.}~\bibnamefont{Dehghani}},
  \bibinfo{author}{\bibfnamefont{T.}~\bibnamefont{Oka}}, \bibnamefont{and}
  \bibinfo{author}{\bibfnamefont{A.}~\bibnamefont{Mitra}},
  \bibinfo{journal}{Phys. Rev. B} \textbf{\bibinfo{volume}{91}},
  \bibinfo{pages}{155422} (\bibinfo{year}{2015}).

\bibitem[{\citenamefont{Halperin}(1982)}]{Halperin1982}
\bibinfo{author}{\bibfnamefont{B.~I.} \bibnamefont{Halperin}},
  \bibinfo{journal}{Phys. Rev. B} \textbf{\bibinfo{volume}{25}},
  \bibinfo{pages}{2185} (\bibinfo{year}{1982}).

\bibitem[{\citenamefont{Onoda et~al.}(2007)\citenamefont{Onoda, Avishai, and
  Nagaosa}}]{OnodaPRL2007}
\bibinfo{author}{\bibfnamefont{M.}~\bibnamefont{Onoda}},
  \bibinfo{author}{\bibfnamefont{Y.}~\bibnamefont{Avishai}}, \bibnamefont{and}
  \bibinfo{author}{\bibfnamefont{N.}~\bibnamefont{Nagaosa}},
  \bibinfo{journal}{Phys. Rev. Lett.} \textbf{\bibinfo{volume}{98}},
  \bibinfo{pages}{076802} (\bibinfo{year}{2007}).

\bibitem[{\citenamefont{Priest et~al.}(2014)\citenamefont{Priest, Lim, and
  Sheng}}]{ShengPRB2014}
\bibinfo{author}{\bibfnamefont{J.}~\bibnamefont{Priest}},
  \bibinfo{author}{\bibfnamefont{S.~P.} \bibnamefont{Lim}}, \bibnamefont{and}
  \bibinfo{author}{\bibfnamefont{D.~N.} \bibnamefont{Sheng}},
  \bibinfo{journal}{Phys. Rev. B} \textbf{\bibinfo{volume}{89}},
  \bibinfo{pages}{165422} (\bibinfo{year}{2014}).

\bibitem[{\citenamefont{Obuse et~al.}(2008)\citenamefont{Obuse, Furusaki, Ryu,
  and Mudry}}]{Obuse2008}
\bibinfo{author}{\bibfnamefont{H.}~\bibnamefont{Obuse}},
  \bibinfo{author}{\bibfnamefont{A.}~\bibnamefont{Furusaki}},
  \bibinfo{author}{\bibfnamefont{S.}~\bibnamefont{Ryu}}, \bibnamefont{and}
  \bibinfo{author}{\bibfnamefont{C.}~\bibnamefont{Mudry}},
  \bibinfo{journal}{Phys. Rev. B} \textbf{\bibinfo{volume}{78}},
  \bibinfo{pages}{115301} (\bibinfo{year}{2008}).

\bibitem[{\citenamefont{Kramer and MacKinnon}(1993)}]{KramerMacKinnon1993}
\bibinfo{author}{\bibfnamefont{B.}~\bibnamefont{Kramer}} \bibnamefont{and}
  \bibinfo{author}{\bibfnamefont{A.}~\bibnamefont{MacKinnon}},
  \bibinfo{journal}{Reports on Progress in Physics}
  \textbf{\bibinfo{volume}{56}}, \bibinfo{pages}{1469} (\bibinfo{year}{1993}).

\bibitem[{\citenamefont{Evers and Mirlin}(2008)}]{Mirlin2008}
\bibinfo{author}{\bibfnamefont{F.}~\bibnamefont{Evers}} \bibnamefont{and}
  \bibinfo{author}{\bibfnamefont{A.~D.} \bibnamefont{Mirlin}},
  \bibinfo{journal}{Rev. Mod. Phys.} \textbf{\bibinfo{volume}{80}},
  \bibinfo{pages}{1355} (\bibinfo{year}{2008}).

\bibitem[{\citenamefont{Li et~al.}(2009)\citenamefont{Li, Chu, Jain, and
  Shen}}]{LiTAIPRL2009}
\bibinfo{author}{\bibfnamefont{J.}~\bibnamefont{Li}},
  \bibinfo{author}{\bibfnamefont{R.-L.} \bibnamefont{Chu}},
  \bibinfo{author}{\bibfnamefont{J.~K.} \bibnamefont{Jain}}, \bibnamefont{and}
  \bibinfo{author}{\bibfnamefont{S.-Q.} \bibnamefont{Shen}},
  \bibinfo{journal}{Phys. Rev. Lett.} \textbf{\bibinfo{volume}{102}},
  \bibinfo{pages}{136806} (\bibinfo{year}{2009}).

\bibitem[{\citenamefont{Groth et~al.}(2009)\citenamefont{Groth, Wimmer,
  Akhmerov, Tworzyd\l{}o, and Beenakker}}]{BeenakkerTAI2009}
\bibinfo{author}{\bibfnamefont{C.~W.} \bibnamefont{Groth}},
  \bibinfo{author}{\bibfnamefont{M.}~\bibnamefont{Wimmer}},
  \bibinfo{author}{\bibfnamefont{A.~R.} \bibnamefont{Akhmerov}},
  \bibinfo{author}{\bibfnamefont{J.}~\bibnamefont{Tworzyd\l{}o}},
  \bibnamefont{and} \bibinfo{author}{\bibfnamefont{C.~W.~J.}
  \bibnamefont{Beenakker}}, \bibinfo{journal}{Phys. Rev. Lett.}
  \textbf{\bibinfo{volume}{103}}, \bibinfo{pages}{196805}
  (\bibinfo{year}{2009}).

\bibitem[{\citenamefont{Xing et~al.}(2011)\citenamefont{Xing, Zhang, and
  Wang}}]{XingHaldaneTAI2011}
\bibinfo{author}{\bibfnamefont{Y.}~\bibnamefont{Xing}},
  \bibinfo{author}{\bibfnamefont{L.}~\bibnamefont{Zhang}}, \bibnamefont{and}
  \bibinfo{author}{\bibfnamefont{J.}~\bibnamefont{Wang}},
  \bibinfo{journal}{Phys. Rev. B} \textbf{\bibinfo{volume}{84}},
  \bibinfo{pages}{035110} (\bibinfo{year}{2011}).

\bibitem[{\citenamefont{Guo et~al.}(2010)\citenamefont{Guo, Rosenberg, Refael,
  and Franz}}]{RefaelTAI3D2010}
\bibinfo{author}{\bibfnamefont{H.-M.} \bibnamefont{Guo}},
  \bibinfo{author}{\bibfnamefont{G.}~\bibnamefont{Rosenberg}},
  \bibinfo{author}{\bibfnamefont{G.}~\bibnamefont{Refael}}, \bibnamefont{and}
  \bibinfo{author}{\bibfnamefont{M.}~\bibnamefont{Franz}},
  \bibinfo{journal}{Phys. Rev. Lett.} \textbf{\bibinfo{volume}{105}},
  \bibinfo{pages}{216601} (\bibinfo{year}{2010}).

\bibitem[{\citenamefont{Yao et~al.}(2017)\citenamefont{Yao, Potter, Potirniche,
  and Vishwanath}}]{Yao_2016}
\bibinfo{author}{\bibfnamefont{N.~Y.} \bibnamefont{Yao}},
  \bibinfo{author}{\bibfnamefont{A.~C.} \bibnamefont{Potter}},
  \bibinfo{author}{\bibfnamefont{I.-D.} \bibnamefont{Potirniche}},
  \bibnamefont{and}
  \bibinfo{author}{\bibfnamefont{A.}~\bibnamefont{Vishwanath}},
  \bibinfo{journal}{Phys. Rev. Lett.} \textbf{\bibinfo{volume}{118}},
  \bibinfo{pages}{030401} (\bibinfo{year}{2017}).

\bibitem[{\citenamefont{Zhang et~al.}(2016)\citenamefont{Zhang, Hess,
  Kyprianidis, Becker, Lee, Smith, Pagano, Potirniche, Potter, Vishwanath
  et~al.}}]{Monroe_2016}
\bibinfo{author}{\bibfnamefont{J.}~\bibnamefont{Zhang}},
  \bibinfo{author}{\bibfnamefont{P.~W.} \bibnamefont{Hess}},
  \bibinfo{author}{\bibfnamefont{A.}~\bibnamefont{Kyprianidis}},
  \bibinfo{author}{\bibfnamefont{P.}~\bibnamefont{Becker}},
  \bibinfo{author}{\bibfnamefont{A.}~\bibnamefont{Lee}},
  \bibinfo{author}{\bibfnamefont{J.}~\bibnamefont{Smith}},
  \bibinfo{author}{\bibfnamefont{G.}~\bibnamefont{Pagano}},
  \bibinfo{author}{\bibfnamefont{I.-D.} \bibnamefont{Potirniche}},
  \bibinfo{author}{\bibfnamefont{A.~C.} \bibnamefont{Potter}},
  \bibinfo{author}{\bibfnamefont{A.}~\bibnamefont{Vishwanath}},
  \bibnamefont{et~al.}, \bibinfo{journal}{arXiv:1609.08684}
  (\bibinfo{year}{2016}).

\bibitem[{\citenamefont{Potirniche et~al.}(2016)\citenamefont{Potirniche,
  Potter, Schleier-Smith, Vishwanath, and Yao}}]{Potirniche_2016}
\bibinfo{author}{\bibfnamefont{I.-D.} \bibnamefont{Potirniche}},
  \bibinfo{author}{\bibfnamefont{A.~C.} \bibnamefont{Potter}},
  \bibinfo{author}{\bibfnamefont{M.}~\bibnamefont{Schleier-Smith}},
  \bibinfo{author}{\bibfnamefont{A.}~\bibnamefont{Vishwanath}},
  \bibnamefont{and} \bibinfo{author}{\bibfnamefont{N.~Y.} \bibnamefont{Yao}},
  \bibinfo{journal}{arXiv:1610.07611}  (\bibinfo{year}{2016}).

\bibitem[{\citenamefont{Roy and Sreejith}(2016)}]{Sthitadhi_PRB_2016}
\bibinfo{author}{\bibfnamefont{S.}~\bibnamefont{Roy}} \bibnamefont{and}
  \bibinfo{author}{\bibfnamefont{G.~J.} \bibnamefont{Sreejith}},
  \bibinfo{journal}{Phys. Rev. B} \textbf{\bibinfo{volume}{94}},
  \bibinfo{pages}{214203} (\bibinfo{year}{2016}).

\bibitem[{\citenamefont{Gannot}(2015)}]{Gannot_arxiv2015}
\bibinfo{author}{\bibfnamefont{Y.}~\bibnamefont{Gannot}},
  \bibinfo{journal}{arXiv:1512.04190}  (\bibinfo{year}{2015}).

\bibitem[{\citenamefont{Titum et~al.}(2015)\citenamefont{Titum, Lindner,
  Rechtsman, and Refael}}]{TitumPRL2015}
\bibinfo{author}{\bibfnamefont{P.}~\bibnamefont{Titum}},
  \bibinfo{author}{\bibfnamefont{N.~H.} \bibnamefont{Lindner}},
  \bibinfo{author}{\bibfnamefont{M.~C.} \bibnamefont{Rechtsman}},
  \bibnamefont{and} \bibinfo{author}{\bibfnamefont{G.}~\bibnamefont{Refael}},
  \bibinfo{journal}{Phys. Rev. Lett.} \textbf{\bibinfo{volume}{114}},
  \bibinfo{pages}{056801} (\bibinfo{year}{2015}).

\bibitem[{\citenamefont{Bernevig et~al.}(2006)\citenamefont{Bernevig, Hughes,
  and Zhang}}]{BHZ}
\bibinfo{author}{\bibfnamefont{B.~A.} \bibnamefont{Bernevig}},
  \bibinfo{author}{\bibfnamefont{T.~L.} \bibnamefont{Hughes}},
  \bibnamefont{and} \bibinfo{author}{\bibfnamefont{S.-C.} \bibnamefont{Zhang}},
  \bibinfo{journal}{Science} \textbf{\bibinfo{volume}{314}},
  \bibinfo{pages}{1757} (\bibinfo{year}{2006}).

\bibitem[{\citenamefont{Loring and Hastings}(2010)}]{hastingsLoringEPL2010}
\bibinfo{author}{\bibfnamefont{T.~A.} \bibnamefont{Loring}} \bibnamefont{and}
  \bibinfo{author}{\bibfnamefont{M.~B.} \bibnamefont{Hastings}},
  \bibinfo{journal}{EPL (Europhysics Letters)} \textbf{\bibinfo{volume}{92}},
  \bibinfo{pages}{67004} (\bibinfo{year}{2010}).

\bibitem[{\citenamefont{Hastings and Loring}(2010)}]{HastingsLoringJmathPhys}
\bibinfo{author}{\bibfnamefont{M.~B.} \bibnamefont{Hastings}} \bibnamefont{and}
  \bibinfo{author}{\bibfnamefont{T.~A.} \bibnamefont{Loring}},
  \bibinfo{journal}{Journal of Mathematical Physics}
  \textbf{\bibinfo{volume}{51}}, \bibinfo{eid}{015214} (\bibinfo{year}{2010}).

\bibitem[{\citenamefont{Avron and Seiler}(1985)}]{Avron1985}
\bibinfo{author}{\bibfnamefont{J.~E.} \bibnamefont{Avron}} \bibnamefont{and}
  \bibinfo{author}{\bibfnamefont{R.}~\bibnamefont{Seiler}},
  \bibinfo{journal}{Phys. Rev. Lett.} \textbf{\bibinfo{volume}{54}},
  \bibinfo{pages}{259} (\bibinfo{year}{1985}).

\bibitem[{\citenamefont{Shklovskii et~al.}(1993)\citenamefont{Shklovskii,
  Shapiro, Sears, Lambrianides, and Shore}}]{Shklovskii_PRB_1993}
\bibinfo{author}{\bibfnamefont{B.~I.} \bibnamefont{Shklovskii}},
  \bibinfo{author}{\bibfnamefont{B.}~\bibnamefont{Shapiro}},
  \bibinfo{author}{\bibfnamefont{B.~R.} \bibnamefont{Sears}},
  \bibinfo{author}{\bibfnamefont{P.}~\bibnamefont{Lambrianides}},
  \bibnamefont{and} \bibinfo{author}{\bibfnamefont{H.~B.} \bibnamefont{Shore}},
  \bibinfo{journal}{Phys. Rev. B} \textbf{\bibinfo{volume}{47}},
  \bibinfo{pages}{11487} (\bibinfo{year}{1993}).

\bibitem[{\citenamefont{Mehta}(2004)}]{Mehta2004}
\bibinfo{author}{\bibfnamefont{M.}~\bibnamefont{Mehta}},
  \emph{\bibinfo{title}{Random Matrices, Pure and Applied Mathematics}}, vol.
  \bibinfo{volume}{142} (\bibinfo{publisher}{Elsevier/Academic Press},
  \bibinfo{address}{Amsterdam, Netherlands}, \bibinfo{year}{2004}),
  \bibinfo{edition}{3rd} ed.

\bibitem[{\citenamefont{D'Alessio and Rigol}(2014)}]{DAlessio2014}
\bibinfo{author}{\bibfnamefont{L.}~\bibnamefont{D'Alessio}} \bibnamefont{and}
  \bibinfo{author}{\bibfnamefont{M.}~\bibnamefont{Rigol}},
  \bibinfo{journal}{Phys. Rev. X} \textbf{\bibinfo{volume}{4}},
  \bibinfo{pages}{041048} (\bibinfo{year}{2014}).

\bibitem[{\citenamefont{Haldane}(1988)}]{Haldanemod1988}
\bibinfo{author}{\bibfnamefont{F.~D.~M.} \bibnamefont{Haldane}},
  \bibinfo{journal}{Phys. Rev. Lett.} \textbf{\bibinfo{volume}{61}},
  \bibinfo{pages}{2015} (\bibinfo{year}{1988}).

\bibitem[{\citenamefont{Martinez}(2003)}]{Martinez}
\bibinfo{author}{\bibfnamefont{D.~F.} \bibnamefont{Martinez}},
  \bibinfo{journal}{Journal of Physics A: Mathematical and General}
  \textbf{\bibinfo{volume}{36}}, \bibinfo{pages}{9827} (\bibinfo{year}{2003}).

\end{thebibliography}
\end{document}